\documentclass[twocolumn,secnumarabic,amssymb,nobibnotes,aps,showpacs,nofootinbib,noeprint]{revtex4-1}
\usepackage[english]{babel}
\usepackage{graphicx}

\usepackage{amsmath}
\usepackage{amsfonts}
\usepackage[usenames,dvipsnames]{color}
\usepackage{hyperref}
\usepackage{rotating}
\usepackage{booktabs} 
\usepackage{slashed}
\usepackage{transparent}

\DeclareMathAlphabet{\mathbb}{U}{bbold}{m}{n}

\begin{document}
\title{Dense ionic fluids confined in planar capacitors: In- and out-of-plane structure from classical density functional theory}
\author{Andreas H\"{a}rtel,$^{1,2}$ Sela Samin,$^2$ and Ren\'{e} van Roij$^2$} 
\affiliation{$^1$ Institute of Physics, 
Johannes Gutenberg-University Mainz, 
Staudinger Weg 9, 55128 Mainz, Germany\\
$^2$ Institute for Theoretical Physics, Center for Extreme Matter and Emergent Phenomena, 
Utrecht University, Leuvenlaan 4, 3584 CE Utrecht, The Netherlands}
\date{\today}

\begin{abstract}
The ongoing scientific interest in the properties and structure of 
electric double layers (EDLs) stems from their pivotal role in 
(super)capacitive energy storage, energy harvesting, and water 
treatment technologies. Classical density functional theory (DFT) is 
a promising framework for the study of the in- and out-of-plane structural properties 
of double layers. 
Supported by molecular dynamics simulations, we 
demonstrate the adequate performance of DFT for analyzing charge layering 
in the EDL perpendicular to the electrodes. We discuss 
charge storage and capacitance of the EDL and the impact 
of screening due to dielectric solvents. 
We further calculate, for the first time, 
the in-plane structure of the EDL within the framework of DFT. While our out-of-plane results 
already hint at structural in-plane transitions inside the EDL, 
which have been observed recently in simulations and experiments, our 
DFT approach performs poorly in predicting in-plane structure in comparison to 
simulations. However, our findings isolate 
fundamental issues in the theoretical description of the EDL within the primitive 
model and point towards limitations in the performance 
of DFT in describing the out-of-plane structure of the EDL at 
high concentrations and potentials. 
\end{abstract}

\maketitle

\section{Introduction}

Nowadays, the efficient storage of electric charges is more important than ever due to 
our continuously increasing need for electric energy. 
For this reason, (super)capacitors 
have attracted large interest in recent years
\cite{simon_nm7_2008,vatamanu_jpcl6_2015}. 
Their electrodes 
\cite{simon_nm7_2008,zhu_science332_2011} 
can be utilized for the construction of sustainable 
energy-conversion 
\cite{brogioli_prl103_2009,boon_mp109_2011,rica_entropy15_2013, 
hamelers_estl1_2014,haertel_ees8_2015} 
and capacitive deionization \cite{porada_ees6_2013,suss_ees8_2015} technology. 
All these devices exploit the capacitive properties of the 
electric double layers (EDLs), which are established in the vicinity of 
electrode surfaces by ionic charges, drawn from the ionic electrolytes 
the electrodes are immersed in \cite{rogers_science302_2003}. 
Accordingly, many studies have focused on the detailed description of the EDL and 
its properties during the last years 
\cite{oleksy_mp104_2006,
kornyshev2007double,fedorov_jpcb112_2008,georgi_jec649_2010,jiang_cpl504_2011,
henderson_jpcb116_2012,lamperski_jcp139_2013,breitsprecher_jpcm26_2014,han_jpcm26_2014,
kong_jcis449_2015}. 
%
%
%
%
%
%

In a recent study, Merlet and co-workers proposed an explanation for some non-linear response in 
the differential capacitance of a parallel plate capacitor by analyzing the in-plane structure of the 
EDL by means of molecular dynamics (MD) simulations 
\cite{merlet_ea101_2013,merlet_jpcc118_2014}. 
They found a voltage-dependent structural transition in the first ionic layer of a 
common ionic liquid (BMI-PF$_6$) in contact with the electrodes, 
which they constructed from carbon particles. Further simulation studies have verified hints for this 
voltage-dependent structural transition 
\cite{kirchner_ea110_2013,breitsprecher_jpcm26_2014,rotenberg_jpcl6_2015} 
and, recently, also experimental studies on the structure of EDLs have been performed 
\cite{jeon_prl108_2012,griffin_nm14_2015}. However, a detailed theoretical 
description beyond these and older work on interfaces 
\cite{baus_molphys48_1983} is still missing. 
%
%
%

A promising microscopic theory, based on fundamental statistical physics, is classical density 
functional theory (DFT). Originally introduced for electronic systems in 1964 
\cite{hohenberg_pr136_1964,jones_rmp87_2015}, 
the framework has also been adopted and applied to classical systems 
\cite{mermin_pr137_1965,ebner_pra14_1976,evans_ap28_1979}, 
especially in the field of soft condensed matter 
\cite{tarazona_inbook_2008,wu_aiche52_2006,hansen_book_2013}, but also in 
(commercial) tools for gas sorption data analysis of porous materials 
\cite{seaton_carbon27_1989,thommes_inbook_2004,ravikovitch_langmuir22_2006}. 
The DFT framework of fundamental measure theory (FMT)
\cite{rosenfeld_prl63_1989,roth_jpcm22_2010,marechal_pre90_2014} has been shown 
to provide a quantitative benchmark for the important model system of hard 
spheres \cite{oettel_pre86_2012}, where it resolved the long-standing 
question on the interfacial free energy in the crystal-fluid 
interface \cite{haertel_prl108_2012}. Even more, FMT accurately predicts pair 
correlations in confined, dense, and asymmetric, mixtures of hard spheres \cite{haertel_pre92_2015}. 

In contrast, a comparably successful functional for the primitive model of {\it charged} (asymmetric) 
hard spheres 
is still missing. While charges have been incorporated into DFT in several forms 
\cite{alts_cp111_1987,mier-y-teran_jcp92_1990,patra_pre47_1993,biben_pre57_1998,
gillespie_pre68_2003,forsman_jpcb115_2011,wang_jpcm23_2011,henderson_jpcb116_2012,hansen_book_2013}, 
their description typically lacks the correct treatment of the interplay between Coulombic and steric 
contributions, as known from the Mean Spherical Approximation in 
bulk \cite{waisman_jcp52_1970,hoye_jcp61_1974,blum_mp30_1975,stell_jcp63_1975, 
sanchez-diaz_jcp132_2010}. 
Thus, the incorporation of charges into DFT is of ongoing scientific 
interest and involves fundamental issues such as testing the contact value theorem 
\cite{henderson_jecie102_1979,gillespie_pre90_2014} 
and analyzing the decay of correlations 
\cite{ulander_jcp114_2001,evans_jpcm21_2009}, 
but is also of relevance for devices with charged interfaces such as supercapacitors. 
%
%
%
%
%
%
%
%
%
%
%

In this work, we apply DFT in order to investigate the fundamental properties of 
the in- and out-of-plane structure of the EDL. 
To ensure a theoretical description of particle ordering in our primitive model, we apply the 
White Bear mark II fundamental measure functional \cite{hansen-goos_jpcm18_2006}
for the hard-sphere contributions, which has been shown to provide excellent 
pair-correlation functions in uncharged systems \cite{haertel_pre92_2015}. 
Further, we solely add a mean-field Coulombic contribution, as described in 
previous work \cite{haertel_jpcm27_2015}. 
Since we extract pair correlations from DFT for the primitive model for the 
first time, we neglect the additional (approximate) correction discussed in 
previous work \cite{haertel_jpcm27_2015} in order to add only one new 
contribution to the well performing hard sphere functional. 
This procedure seems to be adequate for monovalent ions at room temperature \cite{oleksy_mp104_2006}, 
which we use in our primitive model of binary charged hard spheres, 
but now also at lower dielectric constants, i.e. at stronger Coulomb coupling. 

In the next section, we explain in detail our theoretical framework of DFT and the extraction 
of correlation functions in the primitive model. Then, we discuss our choice of model and parameters, 
as well as the differences between DFT and simulations, in Sec.~\ref{sec:model}. Turning to our results, 
we first apply our theory 
in Sec.~\ref{sec:quantifying} to the out-of-plane order and test it against MD computer 
simulations. We further discuss the layering and adsorption of charges in the EDL, as well 
as its capacitance. Second, we apply our theory in Sec.~\ref{sec:inplane-structure} for 
the first time to the in-plane order of the EDL in the inhomogeneous primitive model. 
We discuss our results in comparison with the simulations and 
conclude in Sec.~\ref{sec:conclusion}.

\section{Structure in the primitive model}
\label{sec:theory}

\subsection{The primitive model}

\begin{figure}
\centering
\includegraphics[width=8.0cm]{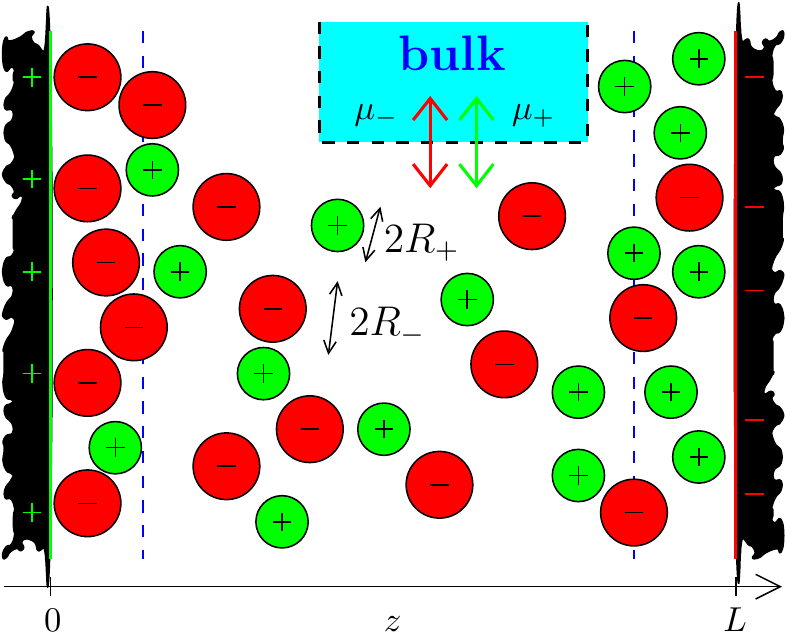}
\caption{\label{fig:sketch}(Color online) 
Sketch of the binary ionic liquid, confined in a parallel plate capacitor of 
plate separation $L$. The stroked lines illustrate the slab of the first layers 
of ions adjacent to the walls, defined by the first minima in the total 
concentration profiles $\rho(z)$. The grand canonical nature of the system is 
explained by the schematic connection to a bulk reservoir
at fixed chemical potentials, which can exchange ions with the system. 
}
\end{figure}

We use the primitive model to describe an ionic liquid. In this model, 
sketched in Fig.~\ref{fig:sketch}, 
the ions of species $\nu$ are described as charged hard spheres with 
hard-sphere radius $R_{\nu}$ and charge valency $e Z_{\nu}$, where $e$ denotes the 
unit charge. Within the primitive model the solvent only enters as a 
dielectric background with relative permittivity $\varepsilon$ and 
temperature $T$. The interaction potential 
$\Phi_{\nu\nu'}$ between two 
particles of species $\nu$ and $\nu'$ with a core separation $r$ reads 
\begin{align}
\Phi_{\nu\nu'}(r) 
= 
\begin{cases}
\displaystyle{\infty } & r<R_\nu+R_{\nu'}; \\
\displaystyle{k_{\rm B}T \lambda_{\rm B} \frac{Z_{\nu} Z_{\nu'}}{r}} & r\geq R_{\nu}+R_{\nu'} , 
\end{cases} \label{eq:interaction-potential}
\end{align}
where $\lambda_{\rm B}=e^2/(4\pi\varepsilon_0\varepsilon k_{\rm B}T)$ 
is the Bjerrum length in terms of 
the vacuum permittivity $\varepsilon_0$ and Boltzmann's constant $k_{\rm B}$. 
In an electrically neutral bulk with mean particle concentrations 
$\bar{\rho}_{\nu}$ for each 
species, this potential is typically screened within a distance characterized 
by the Debye length $\kappa^{-1}$, defined via $\kappa^2=4\pi\lambda_{\rm 
B}\sum_{\nu}Z_{\nu}^2\bar{\rho}_{\nu}$. 

In our grand canonical ensemble (see Fig.~\ref{fig:sketch}) the mean particle 
concentrations 
$\bar{\rho}_{\nu}$ are related to the mean particle numbers $N_{\nu}$ in the 
total system volume $V$ via a spatial integration of the 
ensemble-averaged one-particle concentration (or density) profiles $\rho_{\nu}(\vec{r})$ 
of the respective species $\nu$ at positions $\vec{r}$ 
\cite{hansen_book_2013}. 
Consequently, 
the number of unit charges of species $\nu$ in a partial volume $V'\subset V$ is 
\begin{align}
	Q_{\nu}(V') &= \int_{V'} Z_{\nu}\rho_{\nu}(\vec{r}) d\vec{r} . 
\end{align}
In addition, 
we define the total number of positive and negative unit charges in the total 
system volume $V$ by 
$Q_{+}$ and $Q_{-}$. 

In our confined setting of a parallel plate capacitor, symmetry reduces the 
spatial parameters to only the cartesian $z$ component perpendicular to the hard 
capacitor walls (electrodes). These walls are located at $z=0$ 
and 
$z=L$, such that the concentration profile $\rho_{\nu}(z)$ of a species $\nu$ 
vanishes outside 
the interval $[R_\nu,L-R_\nu]$. This leads to species-dependent Stern layers of thickness 
$R_\nu$ (see also Fig.~\ref{fig:sketch}). 
The wall charges $eQ_{\rm p}$ and $eQ_{\rm n}$ on the respective positive and negative 
electrodes are distributed homogeneously, resulting in wall unit charge densities 
$\sigma_{\rm p}$ and $\sigma_{\rm n}$. Of course, electroneutrality requires 
$Q_{\rm p}+Q_{\rm n}+\sum_\nu Q_\nu=0$. 

All charges in the system contribute to a (dimensionless) electrostatic potential 
$\phi$, which, for the ionic charge density $q(z)=\sum_{\nu}Z_{\nu}\rho_\nu(z)$, 
can be defined via Poisson's equation 
$\partial_{z}^2\phi(z)=-4\pi\lambda_{\rm B}q(z)$ on the open interval $(0,L)$. 
The limits of the Poisson equation, 
\begin{align}
	\lim_{z\to 0^+} \phi'(z) &= -4\pi\lambda_{\rm B} \sigma_{\rm p} , \label{eq:Poisson-limit-anode} \\
	\lim_{z\to L^-} \phi'(z) &= 4\pi\lambda_{\rm B} \sigma_{\rm n} \label{eq:Poisson-limit-cathode}
\end{align}
respect the electrode charges, which are uniquely determined via the ionic charge 
density $q(z)$ and the boundary values 
$\phi(0)=\beta e\Psi_{\rm p}$ and 
$\phi(L)=\beta e\Psi_{\rm n}$ of the electrostatic potential at the capacitor walls, where 
$\beta=(k_{\rm B}T)^{-1}$ defines an inverse temperature. 
Note that these equipotential surfaces, which arise from the system symmetries, set well-defined 
boundaries with no call for additional image-charge methods.

\subsection{Structure and correlations}

In statistical physics the three-dimensional partial static structure factors 
$S_{\nu\nu'}(\vec{k})$ can be defined 
for the total number of particles $N=\sum_{\nu}N_{\nu}$ 
by \cite{hansen_book_2013} 
\begin{align}
S_{\nu\nu'}(\vec{k}) 
&= \frac{N_\nu}{N}\delta_{\nu\nu'} 
\label{eq:structure-factor} \\
 & + \frac{1}{N} \int_V \int_V 
 e^{-\imath\vec{k}\cdot\left(\vec{r}-\vec{r}'\right)} 
 \rho_{\nu}(\vec{r}) \rho_{\nu'}(\vec{r}') 
 h_{\nu\nu'}^{(2)}(\vec{r},\vec{r}') d\vec{r} d\vec{r}' \notag . 
\end{align}
The partial structure factors $S_{\nu\nu'}$ can be combined linearly to 
meaningful structure factors like the particle-particle (NN) structure 
$S_{\rm NN}(\vec{k})=\sum_{\nu}\sum_{\nu'} S_{\nu\nu'}(\vec{k})$, 
the particle-charge (NZ) structure 
$S_{\rm NZ}(\vec{k})=\sum_{\nu}\sum_{\nu'} Z_{\nu'} S_{\nu\nu'}(\vec{k})$, 
and the charge-charge (ZZ) structure 
$S_{\rm ZZ}(\vec{k})=\sum_{\nu}\sum_{\nu'} Z_{\nu}Z_{\nu'} S_{\nu\nu'}(\vec{k})$ 
\cite{hansen_book_2013,ulander_jcp114_2001}. 
%
The last line of Eq.~(\ref{eq:structure-factor}) involves the total correlation functions 
$h_{\nu\nu'}^{(2)}$, which 
are related to the direct correlation functions via the 
Ornstein-Zernike relation \cite{hansen_book_2013} 
\begin{align}
h_{\nu\nu'}^{(2)}(\vec{r},\vec{r}') =& c_{\nu\nu'}^{(2)}(\vec{r},\vec{r}') 
\label{eq:ornstein-zernike-relation} \\
+& \sum_{\nu''=1}^{n} \int_V h_{\nu\nu''}^{(2)}(\vec{r},\vec{r}'') 
\rho_{\nu''}(\vec{r}'') c_{\nu''\nu'}^{(2)}(\vec{r}'',\vec{r}') d\vec{r}'' . \notag 
\end{align}

The static structure factor, as introduced in Eq.~(\ref{eq:structure-factor}), is related 
to the total volume $V$, it is determined on. If we are interested in the structure of only 
the $N^{\rm slab}=\sum_{\nu}N_{\nu}^{\rm slab}$ ions which are located in the slab of length 
$L^{\rm slab}$ next to the wall, we have to restrict our calculations to the corresponding set 
$\Gamma_{\nu}^{\rm slab}$ of particles in the slab. Accordingly, we define the ensemble averaged 
partial in-plane structure factors in this slab by 
\begin{align}
	S_{\nu\nu'}(\vec{q}) &= \left\langle \frac{1}{N^{\rm slab}} 
	\sum_{i\in\Gamma_\nu} \sum_{i'\in\Gamma_\nu'} 
	\exp\left( -\imath \vec{q}\cdot(\vec{r}_{i'}-\vec{r}_i)_{xy} \right) \right\rangle , 
	\label{eq:inplane-structure-factor}
\end{align}
involving the projected two-dimensional in-plane vectors $(\vec{r})_{xy}$ and $\vec{q}$. 
In order to obtain an expression similar to Eq.~(\ref{eq:structure-factor}), 
we have to contract the coordinates of the static structure factor along 
the $z$ direction into the wall plane. We start from an in-plane Fourier 
transform of the total correlation function \cite{haertel_pre92_2015}, 
defined by a Hankel transform via 
\begin{align}
h_{\nu\nu'}^{(2)}(z,z',q) 
&= \int h_{\nu\nu'}^{(2)}(z,z',r) e^{-\imath \vec{q}\cdot(\vec{r})_{xy}} d(\vec{r})_{xy} . 
\label{eq:in-plane-fourier-transform}
\end{align}
Following Tarazona and co-workers \cite{tarazona_mph54_1985}, we then define partial transverse 
structure factors $H_{\nu'\nu}$ by contracting the $z'$ coordinate in the slab of interest via 
\begin{align}
	H_{\nu'\nu}^{\rm slab}(z,q) &= \delta_{\nu\nu'} + \int_{\rm slab} \rho_{\nu'}(z') h_{\nu'\nu}(z',z,q) dz' , 
	\label{eq:transversial-structure-factor}
\end{align}
which, for a one-component bulk fluid, resembles the 
structure factor $S_{\nu'\nu}(k)$. 
Similar to the static structure factor, the partial transverse structure factors can be used 
to construct the transverse particle-particle structure 
$H_{\rm NN}(z,q)= \sum_{\nu'}\sum_{\nu}H_{\nu'\nu}(z,q)$, 
the particle-charge structure $H_{\rm ZN}(z,q) = \sum_{\nu'}\sum_{\nu}Z_{\nu'}H_{\nu'\nu}(z,q)$, 
and the charge-charge structure $H_{\rm ZZ}(z,q) = \sum_{\nu'}\sum_{\nu}Z_{\nu'}Z_{\nu} H_{\nu'\nu}(z,q)$. 
Finally, a second contraction along the remaining $z$ coordinate defines an 
in-plane structure factor 
\begin{align}
	\bar{H}_{\nu'\nu}^{\rm slab}(q) 
	&= \frac{1}{N^{\rm slab}} \int_{\rm slab} \rho_{\nu}(z) H_{\nu'\nu}^{\rm slab}(z,q) dz , 
	\label{eq:transversial-inplane-structure-factor}
\end{align}
which resembles the result from Eq.~(\ref{eq:inplane-structure-factor}).

\subsection{Density functional theory}

We apply the framework of (classical) DFT 
\cite{evans_ap28_1979,hansen_book_2013} to our primitive model. 
DFT deals with ensemble averaged concentration 
(density) profiles $\rho_\nu$ of species $\nu$, which describe the particle distributions 
in a grand canonical ensemble. The equilibrium density profiles $\rho_\nu^{\rm eq}$ minimize the 
grand canonical energy 
functional $\Omega(T,V,\Psi_{\rm p},\Psi_{\rm n},\{\mu_\nu\};[\sigma_{\rm p},\sigma_{\rm n},\{\rho_\nu\}])$, 
which (when minimized with respect to $\rho_{\nu}(\vec{r})$, $\sigma_{\rm p}$, and $\sigma_{\rm n}$) 
gives the grand canonical potential 
of the corresponding physical system with temperature $T$, volume $V$, wall potentials 
$\Psi_{\rm p}$ and $\Psi_{\rm n}$, and chemical potentials $\mu_\nu$. 
Thus, the equilibrium 
density profiles can be determined from 
\begin{align}
\left.\frac{\delta\Omega[\{\rho_\nu\}]}{\delta \rho_\nu(\vec{r})}
\right|_{\left\{\rho_\nu=\rho_{\nu}^{\rm (eq)}\right\}} = 0 \quad \forall \nu . 
\label{eq:functional-minimization}
\end{align}
The minimization with respect to the wall charge densities is guaranteed by construction 
on the basis of Eqs.~(\ref{eq:Poisson-limit-anode}) and (\ref{eq:Poisson-limit-cathode}). 

The grand canonical potential $\Omega$ can be obtained from the Helmholtz free energy $F$ 
via a generalized Legendre transform $\Omega=F-\Psi_{\rm p}Q_{\rm p}-\Psi_{\rm n}Q_{\rm n}-\sum\mu_\nu N_\nu$, 
where the extensive unit charges $Q_{{\rm p/n}}$ and particles 
numbers $N_\nu$ 
are replaced by the intensive electrostatic potentials $\Psi_{{\rm p/n}}$ and chemical potentials $\mu_\nu$. 
It is convenient to split the 
corresponding free energy functional ${\cal F}(T,V,\Psi_{{\rm p/n}},\{N_\nu\};[\sigma_{{\rm p/n}},\{\rho_\nu\}])$ 
into an ideal gas part ${\cal F}^{\rm id}$ and an excess part ${\cal F}^{\rm 
exc}$, where the latter 
includes all particle interactions. 

Within DFT, direct pair-correlation functions $c_{\nu\nu'}^{(2)}$ immediately 
follow from a second functional derivative of the excess free energy \cite{evans_ap28_1979,hansen_book_2013}, 
such that 
\begin{align}
	c_{\nu\nu'}^{(2)}(\vec{r},\vec{r}') 
	&= -\beta\frac{\delta^2 {\cal F}^{\rm exc}}{\delta\rho_{\nu}(\vec{r}) \delta\rho_{\nu'}(\vec{r}')} . 
	\label{eq:direct-correlations}
\end{align}
Then, total correlations are obtained via the Ornstein-Zernike relation in 
Eq.~(\ref{eq:ornstein-zernike-relation}). 
In principle, the total correlations can also be obtained from determining the pair distribution 
function $g_{\nu\nu'}^{(2)}(\vec{r},\vec{r}')$ around a particle fixed at $\vec{r}$, but 
this so-called test-particle route is problematic in situations where long-ranged 
particle interactions are 
present due to the finite boundaries of a numerically treated system (see 
also the discussion in 
\cite{haertel_pre92_2015}). We will therefore follow the compressibility route. 

As in our previous work \cite{haertel_jpcm27_2015}, and according to the 
interaction potential of 
Eq.~(\ref{eq:interaction-potential}), we combine the excess free energy functional from 
a hard-sphere part ${\cal F}^{\rm HS}$ and a Coulombic part ${\cal F}^{\rm C}$. 
For the hard-sphere contribution, we apply the so-called White Bear mark II 
functional \cite{hansen-goos_jpcm18_2006} from fundamental measure theory \cite{rosenfeld_prl63_1989} 
in its tensor version \cite{tarazona_prl84_2000}. 
This functional has been shown to provide quantitative results for free energies \cite{oettel_pre82_2010}, 
phase coexistence \cite{haertel_prl108_2012,oettel_pre86_2012}, 
and pair correlations \cite{haertel_pre92_2015}. For the Coulombic part, 
we add the ionic mean-field contribution 
\begin{align}
	{\cal F}^{\rm C}[\{\rho_{\nu}\}] 
&= k_{\rm B}T \frac{V}{2L} \int_{0}^{L} q(z)\phi(z) dz \nonumber \\
&= k_{\rm B}T \frac{\lambda_{\rm B}}{2} 
\int_V \int_V \frac{q(z) q(z')}{|\vec{r}-\vec{r}'|} d\vec{r} d\vec{r}' . 
\label{eq:coulomb-excess-free-energy}
\end{align}
Obviously, additional contributions could be added to correct for the correlations 
between the pure hard-sphere and Coulombic contributions \cite{haertel_jpcm27_2015}, 
as derived, for example, within the mean spherical approximation in bulk 
\cite{waisman_jcp52_1970,hoye_jcp61_1974,blum_mp30_1975,stell_jcp63_1975,sanchez-diaz_jcp132_2010}. 
For inhomogeneous systems, such corrections are still missing, and, 
since we extract pair correlations from DFT including charges for the 
first time, we neglect any additional and approximate correction in 
order to clearly separate the contribution of the Coulombic contributions 
to the excellently performing hard-sphere functional. 

Similar to the free energy, the direct correlations 
$c_{\nu\nu'}^{(2)} = c_{\nu\nu'}^{\rm HS} + c_{\nu\nu'}^{\rm C}$ 
can be split into correlations due to the hard-sphere and Coulombic part of 
the excess free energy. 
The first contribution, arising from the hard-sphere part, was determined in a
recent 
work \cite{haertel_pre92_2015} by one of us, 
where it was shown to be in excellent agreement with Brownian 
dynamics simulations. 
The Coulombic part immediately follows in analytic form from 
Eqs.~(\ref{eq:direct-correlations}) and (\ref{eq:coulomb-excess-free-energy}) and reads 
\begin{align}
c_{\nu\nu'}^{\rm C}(\vec{r},\vec{r}') 
&= -\beta\frac{\delta^2 {\cal F}_{\rm C}}{\delta\rho_{\nu}(\vec{r}) \delta\rho_{\nu'}(\vec{r}')} 
= - \lambda_{\rm B} \frac{Z_{\nu} Z_{\nu'}}{|\vec{r}-\vec{r}'|} . 
\end{align}
Following Eq.~(\ref{eq:in-plane-fourier-transform}), $c_{\nu\nu'}^{\rm C}$ can also be Fourier 
transformed analytically in the $xy$ plane, resulting in 
\begin{align}
	c_{\nu\nu'}^{\rm C}(z,z',q) 
	&= -\frac{\lambda_{\rm B}}{2\pi} \int\int \frac{Z_\nu Z_{\nu'}}{|\vec{r}-\vec{r}'|} 
	e^{-\imath \vec{q}\cdot(\vec{r}')_{xy}} d(\vec{r}')_{xy} 
	\label{eq:c2-coulomb-fouriertransformed-general} \\
	&= -\lambda_{\rm B} Z_\nu Z_{\nu'} \frac{1}{q} e^{-q|z-z'|}
	\label{eq:c2-coulomb-fouriertransformed}
\end{align}
with $q=|\vec{q}|\neq 0$. 
In the limit of vanishing $\Delta z=z'-z\to 0$, Eq.~(\ref{eq:c2-coulomb-fouriertransformed}) 
becomes $-\lambda_{\rm B} Z_{\nu}Z_{\nu'}/q$. 

Inserting Eq.~(\ref{eq:c2-coulomb-fouriertransformed-general}) and the hard-sphere contributions 
$c_{\nu\nu'}^{\rm HS}(z,z',q)$ into Eq.~(\ref{eq:ornstein-zernike-relation}) leads to 
\begin{align}
	&h_{\nu\nu'}^{(2)}(z,z',q) 
= c_{\nu\nu'}^{(2)}(z,z',q) \\
&+ 2\pi \sum_{\lambda} \int_{-\infty}^{\infty} \rho_{\lambda}(z'') 
h_{\nu\lambda}^{(2)}(z,z'',q) c_{\lambda\nu'}^{(2)}(z'',z',q) dz'' \notag , 
\end{align}
which we then use together with the explicit form Eq.~(\ref{eq:c2-coulomb-fouriertransformed}) 
to calculate the in-plane structure via 
Eq.~(\ref{eq:transversial-structure-factor}). 

%
%
%
%
%
%

\section{Choice of model and parameters}
\label{sec:model}

In order to perform quantitative comparisons, we must first choose the 
model parameters appropriately. 
Since this work has been inspired by the findings of 
Merlet and co-workers \cite{merlet_jpcc115_2011,merlet_jpcc118_2014}, we will 
use the Lennard-Jones diameters of the spherical particles in their work as the hard-sphere 
diameter of our particles. 
Furthermore, we approximated the shape of their elongated BMI$^{+}$ ion by a sphere of the same 
volume. Consequently, throughout this work we use ionic diameters 
$2R_{+}=0.618294$ nm for the 
BMI$^{+}$ ions and $2R_{-}=0.506$ nm for the PF$_6^{-}$ ions, having a 
diameter ratio $R_{-}/R_{+}\approx 0.818$ and a 
asymmetry $\alpha\approx 0.2919$, which is defined for the ionic volumes $V_+$ 
and $V_-$ via $\alpha=(1-V_{+}/V_{-})/(1+V_{+}/V_{-})$. 
We also use the same valencies, $Z_{\pm}=\pm 0.78$ $e$, the temperature 
$T=400$ K, and 
the wall separation $L=12.32$ nm as in 
\cite{merlet_jpcc115_2011,merlet_jpcc118_2014}.
However, we chose a different relative permittivity and mean 
concentrations different from 
$\bar{\rho}_{\pm}=2.345$ nm$^{-3}$ 
for the following reasons. 

In the equal-size limit of $\alpha\to 0$, the volume-conserving averaged ionic mean radius 
$\bar{R}\approx 0.5677$ nm allows to draw an approximate comparison between the described binary 
system of Merlet and co-workers and the well-studied restricted primitive model of equally-sized 
ions. With a volume fraction $\eta=4\pi\bar{R}^3\bar{\rho}/3\approx 0.15$ 
and an effective dimensionless temperature $T^*=2\bar{R}/\lambda_{\rm B}\approx 0.01$, the system with 
$\varepsilon=1$ would be located in the metastable region of the fluid-solid coexistence 
\cite{vega_jcp119_2003,hynninen_prl96_2006}, which most probably is avoided 
for the binary mixture by the asymmetry of the ion radii. 
Nonetheless, we have chosen to work at a higher effective 
temperature $T^*$ by varying the relative permittivity between 
$\varepsilon=1$ and $30$, focusing on $\varepsilon=10$ 
if not mentioned otherwise. By this choice, our work 
can be related not only to the ionic liquid BMI-PF$_6$, 
which tends to demix from water \cite{chaumont_jpcb109_2005}, but also to other 
ionic electrolytes with higher permittivities, such as diluted ionic liquids or 
tetraethylammonium tetrafluoroborate 
(TEA-BF$_4$) solvated in acetonitrile (ACN) \cite{haertel_ees8_2015}, 
which is often used in supercapacitors. 
Such systems have typical concentrations of about $\bar{\rho}_{\pm}=0.6$ 
nm$^{-3}$ ($1$ M), 
where the electrodes are still screened well and ionic core repulsions are 
important ($\kappa^{-1}\approx 4.6$ nm). We will focus in the work at hand on 
moderate ionic concentrations of around $2$ M. 
%
Note however that we do not aim to give a realistic model 
for one specific ionic liquid but rather aim to bring forward 
the knowledge about microscopic effects involved in the composition of the EDL, 
which all ionic liquids have in common. 

In this work, we test our DFT results against 
MD simulations of charged pseudo hard-sphere particles. Details of
the simulation method can be found in previous work \cite{haertel_jpcm27_2015}. 
The simulations were performed using the
ESPResSo package \cite{espresso}. There are three notable methodological
differences between the simulations and the theory. First, in
the simulation we sample the canonical ensemble and not the grand-canonical
one, second, we assume transverse symmetry in our theory, and third, 
we impose constant charge densities and not voltage on the
electrodes. To match the theory and simulation setups we use as input for
the simulations the particle number densities and surface charges obtained from
the DFT calculations (see for example Table~\ref{tab:cell-data}). 
This is justified by the fact that, first, there is a large bulk-like
region in the capacitor, and second, we find that the simulations reproduce 
accurately the theoretically predicted densities. Furthermore, we performed bulk simulations of
the ionic liquid and found excellent match between the simulations and theory.
Lastly, we also verified that the correct wall potentials are reproduced in the
simulations.

\section{Out-of-plane structure and charging effects}
\label{sec:quantifying}


\begin{figure}
\centering
\includegraphics[width=8.0cm]{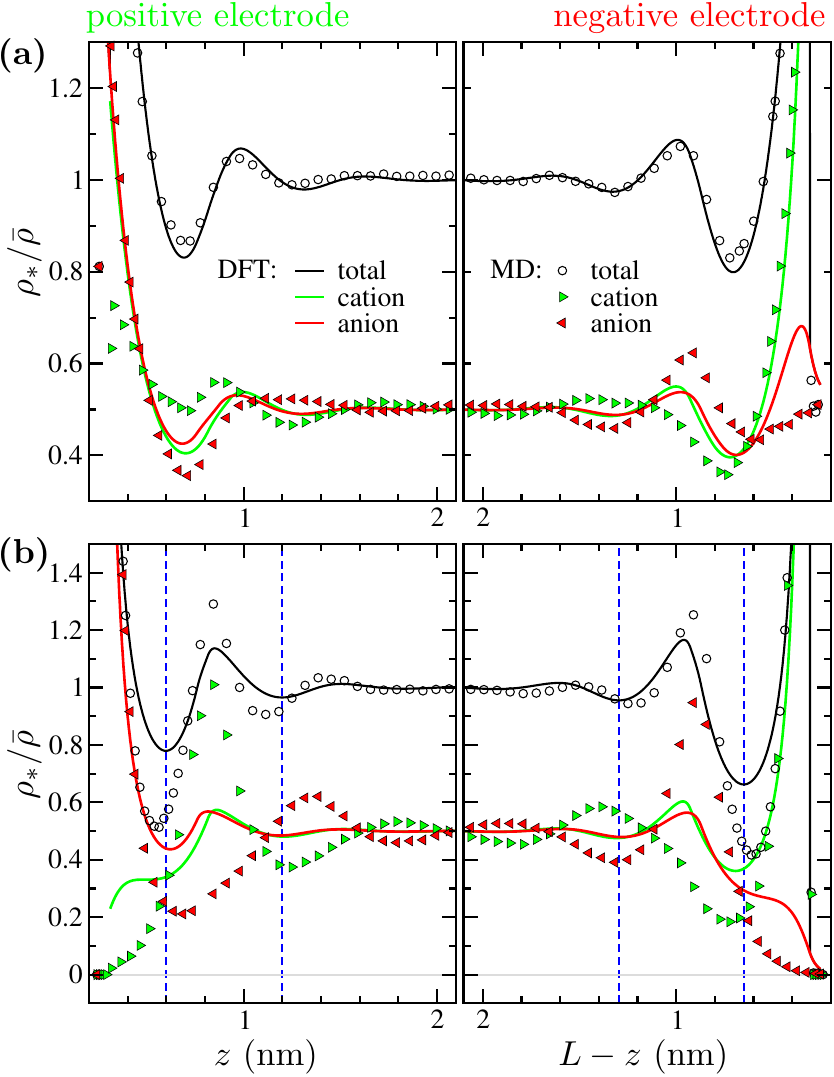}
\caption{\label{fig:cmpsim}(Color online) 
Comparison between DFT results (curves) and MD computer simulations (symbols) 
for the concentration profiles $\rho_*$, where $*$ is a 
place holder for cations, anions, and the sum of both. The profiles are normalized 
to the total concentration $\bar{\rho}=\bar{\rho}_{+}+\bar{\rho}_{-}$ of ions in 
bulk. Results are shown for a system 
with temperature $T=400$ K, permittivity $\varepsilon=10$, and 
bulk concentration $\bar{\rho}_{+/-}=2$ M, at two electrostatic wall potentials 
(a) $\Psi_{\rm p}=-\Psi_{\rm m}=0.1$ V and (b) $0.5$ V. 
The plots show regions in the vicinity of the system walls, which are located 
at $z=0$ nm and $z=L=12.32$ nm. Additionally, 
panel (b) shows the positions of the first two minima in the total 
concentration profiles $\bar{\rho}$ next to the wall, which are marked by vertical 
dashed lines, located at $z=0.598$ nm, $1.198$ nm, $L-1.296$ nm, and $L-0.601$ nm. }
\end{figure}

We first compare the out-of-plane EDL structure resulting from the 
theory and the MD simulations. In Fig.~\ref{fig:cmpsim} we show typical 
ionic concentration profiles for two systems with a reference bulk concentration 
of $2$ M. The electrodes are charged asymmetrically up to potentials (a) $\pm 0.1$ V 
and (b) $\pm 0.5$ V. It is obvious from the profiles that the layering of 
ions in general is captured in our theory even at high packing and high potentials, 
but especially the alternating layering of differently charged ions is clearly 
underestimated. The DFT anion and cation profiles seem to ``stick together'' too strongly 
(see red and green curves in comparison to red and green symbols), while our MD simulations 
and previous work show alternating layers of both species for high potentials and concentrations 
\cite{georgi_jec649_2010,forsman_jpcb115_2011,merlet_jpcc115_2011,lamperski_jcp139_2013,breitsprecher_jpcm26_2014,mezger_jcp142_2015}. 
The reason is most probably that the mean-field description of charges from 
Eq.~(\ref{eq:coulomb-excess-free-energy}) is decoupled from the 
hard-core repulsion \cite{forsman_jpcb115_2011}, 
which overestimates the Coulombic attraction between ionic cores 
at small distances, where particles overlap. 

%
%

\begin{figure}
\centering
\includegraphics[width=8.0cm]{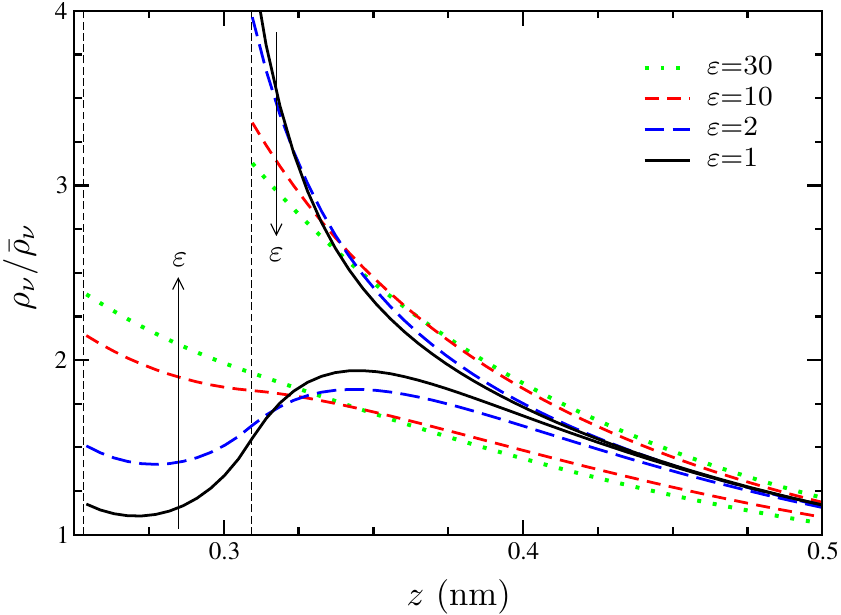}
\caption{\label{fig:profiles}(Color online) 
Concentration profiles of the cations and anions next to the left electrode 
at vanishing potential $\Psi_{\rm p}$ for four relative permittivities 
$\varepsilon=1$, $2$, $10$, and $30$ at an ionic concentration $\bar{\rho}_{+/-}=2$ M. 
Due to their sizes, the anions reach the electrode closer than the larger cations, which leads to a 
separation of the concentration profiles. } 
\end{figure}

The ``sticking effect'' of the ionic profiles described above is also visible 
in Fig.~\ref{fig:profiles}, where we show the profiles for different 
permittivities between uncharged walls. The DFT results 
show the typical layering of a binary system, where the small anions reach the walls 
closer than the larger cations (e.g. \cite{forsman_jpcb115_2011,lamperski_jcp139_2013,breitsprecher_jpcm26_2014}). 
Increasing the dielectric screening by increasing $\varepsilon$ strengthens this layering of anions 
(arrow up in Fig.~\ref{fig:profiles}), while the layering of the cations is weakened 
(arrow down in Fig.~\ref{fig:profiles}). 

\begin{table}
\center
\caption{\label{tab:cell-data} 
Data as obtained from DFT for systems of several relative permittivities $\varepsilon$ at ionic 
concentrations $\bar{\rho}_{+/-}=2$ M. For each system either the electrostatic 
wall potentials $\Psi_{\rm p/n}$ at the positive anode and the negatively charged 
cathode sum up to zero, or the wall charge densities $\sigma_{\rm p/n}$ sum up to 
zero. In addition, the total 
numbers $N_{+/-}$ of ions of each species $+$ and $-$ in the system are given
per unit area of the lateral extension. 
}
\begin{tabular}{ ccccccc }
\toprule[1.5pt]
$\varepsilon$ & $\Psi_{\rm p}$ & $\Psi_{\rm n}$ & $\sigma_{\rm p}$ & $\sigma_{\rm n}$ & $N_{+}$ & $N_{-}$ \\
(1) & (V) & (V) & (e/nm$^{2}$) & (e/nm$^{2}$) & (1/nm$^2$) & (1/nm$^2$) \\
\hline
30 & 0.0 & -0.0 & 0.023 & 0.023 & 14.46 & 14.52 \\
30 & 0.437 & -0.563 & 1.775 & -1.775 & 15.09 & 15.11 \\
10 & 0.0 & -0.0 & 0.019 & 0.019 & 14.46 & 14.51 \\
10 & 0.087 & -0.113 & 0.156 & -0.156 & 14.49 & 14.49 \\
10 & 0.1 & -0.1 & 0.177 & -0.136 & 14.47 & 14.52 \\
10 & 0.459 & -0.541 & 0.777 & -0.777 & 14.67 & 14.68 \\
10 & 0.5 & -0.5 & 0.848 & -0.720 & 14.60 & 14.78 \\
2 & 0.0 & -0.0 & 0.011 & 0.011 & 14.46 & 14.49 \\
1 & 0.0 & -0.0 & 0.008 & 0.008 & 14.45 & 14.48 \\
\midrule
\bottomrule[1.5pt]
\end{tabular}
\end{table}

Indeed, the layer of exclusively anions in contact with the wall introduces a local electric field as 
a counterpart to the mechanical pressure towards the wall. When the potential is forced to 
vanish at the electrodes, electrical charges are induced on them. Thereby, the 
charge induced at the wall rises with increasing permittivity, as shown in 
Table~\ref{tab:cell-data}. Next to the results, where we initially fixed the electrostatic 
potentials $\Psi_{\rm p/n}$ at the electrodes, Table~\ref{tab:cell-data} also shows 
data for fixed surface charge densities $\sigma_{\rm p/n}$ on the 
electrodes. Since the former is natural for our grand canonic theory, we had to scan 
over different potentials to force the charges on both plates to add to zero, which models 
the standard function of a capacitor. Note in this context that we cannot simply shift potentials 
in our system, because the system is in osmotic contact with a neutral bulk reservoir at 
zero potential, i.e. the zero-potential point is already chosen in the bulk reservoir. 

\begin{figure*}
\centering
\includegraphics[width=12.5cm]{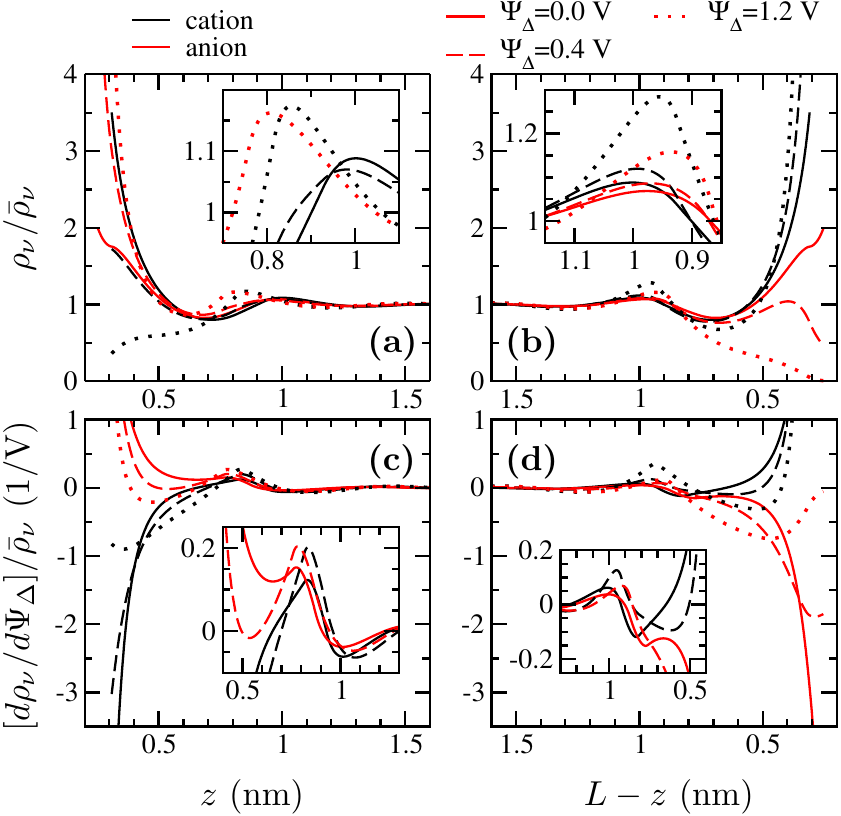}
\caption{\label{fig:profiles-change}(Color online) 
Concentration profiles of cations and anions at (a) the anode and (b) the cathode for 
different cell potentials $\Psi_{\Delta}=\Psi_{\rm p}-\Psi_{\rm n}$ in a system of 
ionic concentrations $\bar{\rho}_{+/-}=2$ M and a relative permittivity $\varepsilon=10$. 
The insets show magnifications of the second peak in the profiles. 
The local change in the ionic concentrations with respect to a change in the cell 
potential is shown in panels (c) and (d), where, again, the insets magnify the regions of 
the second peaks. }
\end{figure*}

In Figs.~\ref{fig:profiles-change} (a) and (b) we show ionic concentration profiles near the 
positive and negative electrode, respectively, for three applied voltages. When the potential difference 
$\Psi_{\Delta}=\Psi_{\rm p}-\Psi_{\rm n}$ between both electrodes is increased, 
the ionic charges, which screen the electrodes, reorganize. Recent work 
indicates that the EDL can be subdivided into 
layers \cite{breitsprecher_jpcm26_2014} with ion exchange 
occurring between them as $\Psi_{\Delta}$ changes. This exchange has also been 
discussed as the cause of structural in-plane transitions within the EDL when 
the potential is increased \cite{merlet_jpcc118_2014}. 
The local change in the ionic concentrations due to the change in $\Psi_{\Delta}$ 
is shown in panels (c) and (d). This response function demonstrates how the layers are 
structured and to which regions ions tend to go to and leave from. For example, 
the magnification in the inset of Fig.~\ref{fig:profiles-change}(c) shows that 
at low potentials the concentration of anions next to the positive electrode increases 
in all regions when the potential is increased (red solid curve positive for all $z<0.9$ nm), 
whereas at higher potentials the change in concentration becomes negative in the region around $z=0.5$ nm, 
corresponding to the minimum around $z=0.6$ nm in the anion concentration profile (red curves) in panel (a). 
Moreover, Fig.~\ref{fig:profiles-change}(d) 
clearly shows how the anions are repelled from the increasingly 
negative electrode: initially, the rejection is strongest immediately at the 
wall (red solid curve in panel (d)), but with increasing potential, when the ionic concentration vanishes at 
the wall (red curves in panel (b)), anions are also rejected from more distant regions where they 
were attracted to earlier (red dotted curve in panel (d)). 

Another salient feature of the EDL that is captured by our DFT is 
the change in the shape of the concentration peaks when the voltage 
increases, which indicates a structural transition in the EDL. The cation 
distributions near the anode in Fig.~\ref{fig:profiles-change}(a) show 
that the first peak of the concentration profile decreases with increase of 
$\Psi_{\Delta}$, vanishing completely for $\Psi_{\Delta}=1.2$ V (black curves). At the same 
time, the peak shifts to larger $z$ and the (initially) second 
peak first decreases (dashed curve) and then builds up again (the black solid curve in panel (a) has 
two peaks, at the wall and at $z\approx 1$ nm, the dotted curve has only one peak at $z\approx 0.8$ nm). 
The final result is that the first peak (black solid curve at the wall) has been shifted by 
$0.6$ nm and combined with the second peak (black dotted curve at $z\approx 0.8$ nm). 
Furthermore, at large voltages the first layer of cations resides 
slightly farther from the electrode compared to the \emph{second} layer of 
anions (dotted curves in the inset of panel (a)). 
These structural features of the EDL are very similar to those described by the 
simulation study of Kirchner \textit{et al.} \cite{kirchner_ea110_2013}, in which it 
is argued that these are additional hallmarks of the \emph{in-plane} ordering of the 
counter-ion layer adsorbed on the wall. To test this within the framework of 
DFT we will investigate the in-plane structure factor of the first adsorbed layer in 
the next section. 


%
%
\begin{figure*}
\centering
\includegraphics[width=13.5cm]{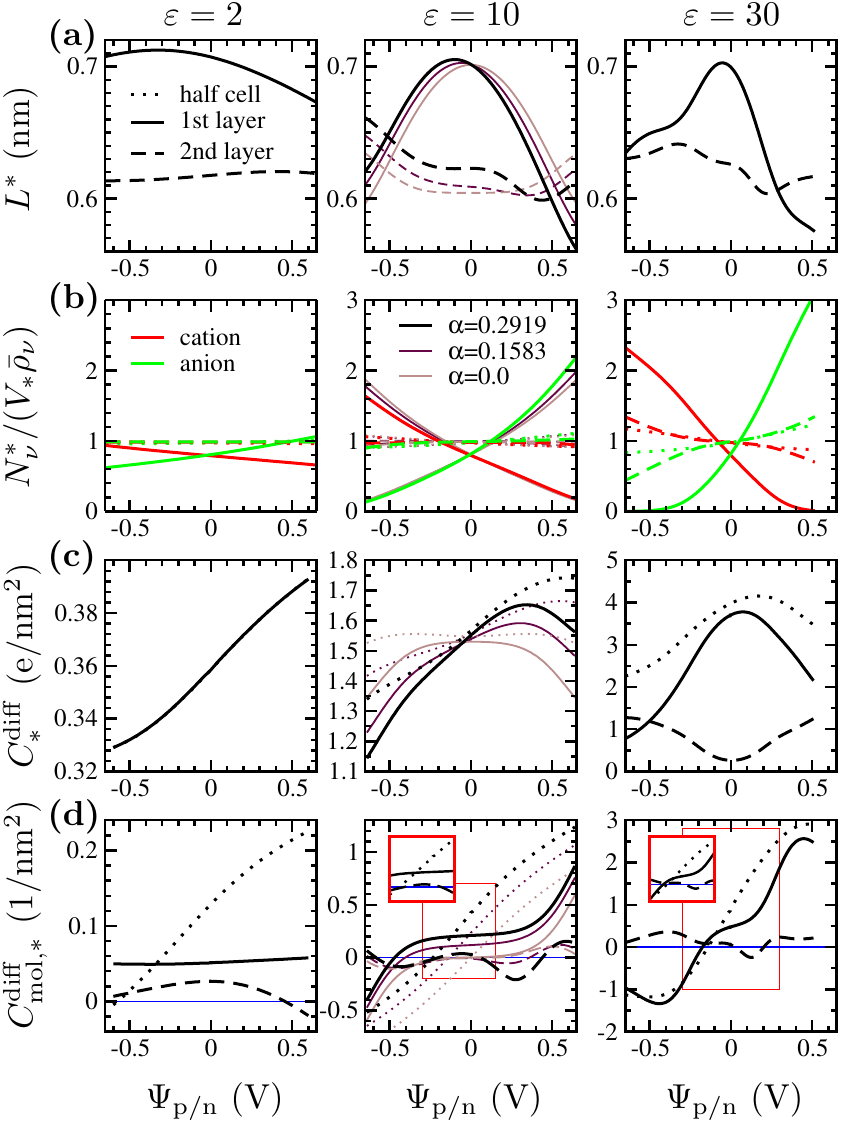}
\caption{\label{fig:charges-in-edl}(Color online) 
Characteristics of the first two layers at the electrode and of the half cell in 
terms of 
(a) their thickness, (b) the amount of particles in relation to bulk, 
(c) the resulting contribution $C_{*}^{\rm diff}$ to the differential capacitance, and 
(d) the contribution $C_{\rm mol,*}^{\rm diff}$ to the molar differential capacitance. 
The results in parts (b), (c), and (d) are separated into cation and anion contributions. 
The three columns show results for the relative permittivities 
$\varepsilon=2$, $\varepsilon=10$, and $\varepsilon=30$, 
each at ionic concentrations of $\bar{\rho}_{+/-}=2$ M. 
In the case of $\varepsilon=10$, data is shown for the additional 
asymmetries $\alpha=0.1583$ and $\alpha=0.0$ of the ionic species. Furthermore, 
magnifications in part (d) demonstrate that a decreasing permittivity seems 
to be similar to a magnification of the graphs at hand, since the magnification 
at $\varepsilon=30$ is qualitatively similar to the plot at $\varepsilon=10$, 
where the magnification again is qualitatively similar to the plot at $\varepsilon=2$. }
\end{figure*}

The layers of the EDL can be defined in between the minima of the total ionic 
concentration profile \cite{breitsprecher_jpcm26_2014}, which we 
show in Fig.~\ref{fig:cmpsim}(b) by the dashed vertical lines. Here, we ignored 
the discontinuous minimum which stems from the fact that the smaller anions can approach 
the wall more closely. The thickness of the layers, which we plot in 
Fig.~\ref{fig:charges-in-edl}(a) for different permittivities, depends on 
the applied electrostatic wall potentials $\Psi_{\rm p/n}$. 
In Fig.~\ref{fig:charges-in-edl}(a), we show the thickness, $L^{*}$, of the first layer, of the 
second layer, and of the half cell ($L/2=6.16$ nm) at the respective electrode, where $*$ just 
represents a place holder. 
Especially at high $\varepsilon$, the thickness of the layers show interesting dependence 
on the applied potential. 
In part (b) of Fig.~\ref{fig:charges-in-edl}, we plot the number $N_{\nu}^{*}$ of 
particles relative to bulk in the respective regions. Obviously, ion exchange 
is most significant in the first layer of ions adjacent to the electrode, 
in accordance with the spatially resolved results in 
Figs.~\ref{fig:profiles-change}(c) and (d). Note, however, that a large 
response in the ionic concentrations does not correspond to a large change in the total number of 
particles, but rather to a large change in the local charge density and, hence, 
in the differential capacitance of the cell (dotted lines remain almost flat in panel (b)). 


The differential capacitances per unit area, shown in part (c) of Fig.~\ref{fig:charges-in-edl},  
are defined as the change of charges per area, $\sigma_*$, in a certain subvolume $V_{*}$ of 
the whole cell due to potentials, hence, 
\begin{align}
	C_{*}^{\rm diff} &= \frac{\partial e\sigma_{*}}{\partial\Psi} . \label{eq:differential-capacitance}
\end{align}
Applied to the whole cell volume, the voltage dependence of 
this quantity is of great importance for the properties 
of charging (super)capacitors, and is therefore often studied in the context of ionic 
liquids \cite{kornyshev2007double,georgi_jec649_2010,jiang_cpl504_2011,
lamperski_jcp139_2013,breitsprecher_jpcm26_2014,han_jpcm26_2014}. 
In particular, Kornyshev was the first to discuss the difference between the 
so-called bell and camel shape of the voltage dependence of the differential capacitance 
\cite{kornyshev2007double}, where our results shown in panel (c) are somewhere in between 
both shapes. For asymmetric particle sizes, these shapes are
asymmetrically deformed for positive and negative electrodes 
\cite{jiang_cpl504_2011,han_jpcm26_2014}, as the 
results in Fig.~\ref{fig:charges-in-edl}(c) confirm for the system with 
$\varepsilon=10$; here, we also provide a comparison between our system's 
particle asymmetry $\alpha\approx 0.2919$ and two less asymmetric 
systems, with a smaller and a vanishing $\alpha$, clearly demonstrating the mentioned 
symmetry for a vanishing $\alpha$ and the asymmetric shift for finite $\alpha$. 
Interestingly, the differential capacitance of the second layer for $\varepsilon=30$ 
seems to be almost symmetric, while the first layer shape shows a large asymmetry. 
Moreover, Fig.~\ref{fig:charges-in-edl}(c) demonstrates that 
the first layer contributes most to the differential capacitance of the respective electrode, 
especially for small permittivities. Indeed, the contribution of the first layer is almost 
identical for our system with $\varepsilon=2$ (the lines are on top of each other). 
Even though the total amount of charge that is stored 
on the electrodes is higher for higher permittivities (see 
Table~\ref{tab:cell-data}), the charges are stored in a much smaller volume for 
low permittivities. For this reason, the permittivity might be decreased in 
(super)capacitors with very narrow pores in order to maximize the ability to 
store electrical charges on the electrodes. 

In contrast to (super)capacitors, where charge capacities on the electrodes are 
crucial, 
molar ion capacities for salt adsorption play an important role in capacitive desalination and 
capacitive energy extraction \cite{rica_prl109_2012,suss_ees8_2015,haertel_jpcm27_2015}. 
The molar differential capacitance 
at fixed chemical potential 
is defined similarly to Eq.~(\ref{eq:differential-capacitance}) by 
\begin{align}
	C_{\rm mol,*}^{\rm diff} &= \frac{\partial {\cal N}_*}{\partial\Psi} , 
\end{align}
where ${\cal N}_{*}$ denotes the total number of particles per electrode area in the volume 
$V_{*}$ of interest. We 
plot this measure of particle storage in Fig.~\ref{fig:charges-in-edl}(d). 
Obviously, $C_{\rm mol,*}^{\rm diff}$ reaches higher values for higher 
permittivities, where the repulsion between like-charge ions is better 
screened. Since the molar differential capacitance is a derivative of the 
particle number, a vanishing value indicates a minimum in the latter. Interestingly, this 
minimum is shifted towards negative potentials for increasing asymmetry $\alpha$, 
such that at the negative electrode the total number of particles is decreased for increasing 
potentials. This can be understood from Fig.~\ref{fig:profiles-change}, which shows that 
the small anions are repelled from the cathode at lower absolute potentials than the cations 
from the anode. In a simple picture, each cation, adsorbed to the cathode, needs 
$(1+\alpha)/(1-\alpha)$ anions to leave in order to have enough free volume, 
as follows from the definition of $\alpha$ in Sec.~\ref{sec:quantifying}. 
As a consequence, small counterions seem to adsorb better to an electrode than small coions. 
In conclusion, the effective adsorption of ions requires a detailed study of the 
adsorption as a function of the ionic sizes and the solvent's permittivity.

\section{In-plane structure of the electric double layer}
\label{sec:inplane-structure}

In this section, we analyze the partial transverse structure factors $S_{\nu'\nu}^{\rm slab}(q)$ 
in the first layer next to the anode. Recall that the first layer is defined 
up to the first minimum of the total concentration profile, $z_{\rm min}$, 
ignoring the minimum at $z=R_{+}$ due to the small anions reaching 
closer to the walls. Naturally, $z_{\rm min}$ depends on the potential 
$\Psi_{\rm p}$, as the analysis of the thickness of layers in Fig.~\ref{fig:charges-in-edl}(a) shows. 
Accordingly, we have calculated $H_{\nu'\nu}^{\rm slab}(z_{0},q)$ from DFT on the 
slab interval $[R_{-},z_{\rm min}]$, as explained in Eq.~(\ref{eq:transversial-structure-factor}). 
For technical reasons and in order to compare our DFT-generated in-plane structure to that obtained from 
our simulations, we have approximated the contraction from 
Eq.~(\ref{eq:transversial-inplane-structure-factor}) by 
$\bar{H}_{\nu'\nu}^{\rm slab}(q)=n_{\nu'}^{\rm slab}H_{\nu'\nu}^{\rm slab}(z_0,q)$
with the mole fraction $n_{\nu'}^{\rm slab}=N_{\nu'}^{\rm slab}/N^{\rm slab}$ in the first layer, 
which is equivalent to $S_{\nu'\nu}^{\rm slab}(q)$. 
As a representative position we have chosen $z_0=(R_{+}+z_{\rm min})/2$, such that the 
slab contains both anions and cations. We did not find any noteworthy differences 
for other values of $z_0$ within the slab. 
%
%
Note that the structure $H_{\nu\nu'}^{\rm slab}(z_0,q)$ still is determined from the complete 
framework of DFT and the Ornstein-Zernike relation applied to the full inhomogeneous, $z$-dependent 
density profiles we discussed in this work. 

\begin{figure}
\centering
\includegraphics[width=8.0cm]{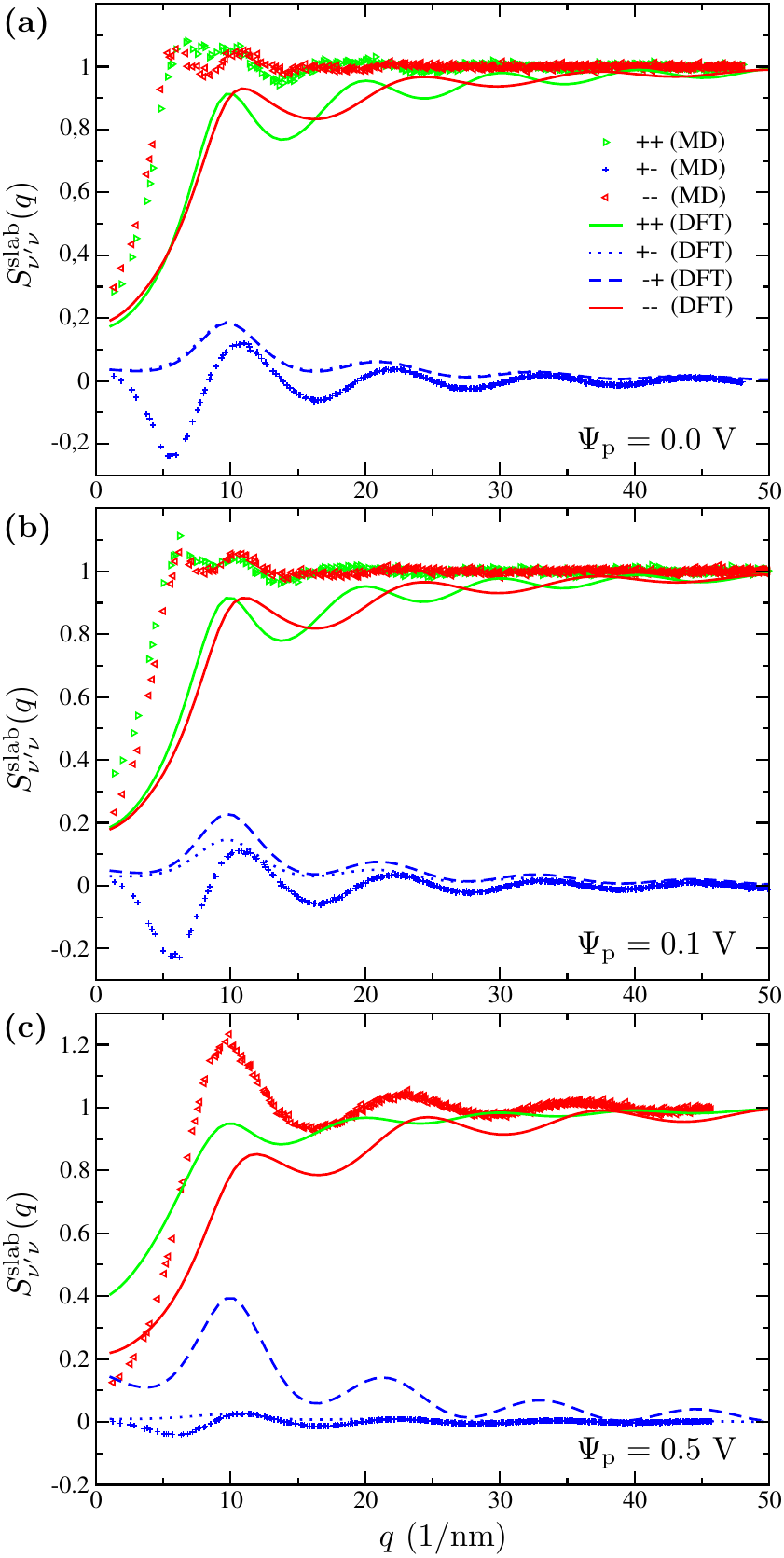}
\caption{\label{fig:inplane-partial}(Color online) 
The partial transverse structure factors $S_{\nu'\nu}^{\rm slab}(q)$ in the first layer 
of ions next to the anode, obtained from both DFT (approximatively) and MD simulations, 
for electrode potentials 
(a) $\Psi_{\rm p}=0.0$ V, (b) $\Psi_{\rm p}=0.1$ V, and (c) $\Psi_{\rm p}=0.5$ V. 
In (c), the $++$ contribution from the MD simulations is not shown 
due to an almost vanishing number of cations, which leads to bad statistics. 
Note that the diagonal parts ($++$ and $--$) of the structure factor have been normalized by the 
mole fraction of the corresponding species to reach comparable large-$q$ limits of $1$. 
}
\end{figure}

In Fig.~\ref{fig:inplane-partial}, we draw a comparison between both the 
partial transverse structure factors $S_{\nu'\nu}^{\rm slab}(q)$, approximatively calculated 
from $n_{\nu'}^{\rm slab}H_{\nu'\nu}^{\rm slab}(z_0,\circ)$ and DFT, and 
obtained from MD simulations, for three electrode potentials. 
The first global observation is that there is certainly no quantitative 
agreement between the DFT results and the simulations. Such a poor agreement 
should not come as a surprise given (i) the incomplete construction of the free-energy functional 
by neglecting correlations between the hard-sphere and electrostatic contributions and 
(ii) the resulting issues in the out-of-plane structure we discussed in the previous section. 
In particular, we observe 
in Fig.~\ref{fig:inplane-partial} that the simulated in-plane structure is more 
pronounced for the diagonal ($++$ and $--$) components, reaching the asymptotic 
large-$q$ limit for smaller values of $q$. This is consistent with the 
``sticking'' effect of our functional, which gives spurious structure on small 
length scales and hence on large $q$'s. The slow large-$q$ decay of the 
DFT-based structure may well be due to this shortcoming of the DFT. 
Figure~\ref{fig:inplane-partial} also shows a relatively large difference between 
panels (a) and (b) compared to panel (c), where the $--$ correlations obtained 
by the simulations become very pronounced at $q\sim10$ nm$^{-1}$, although this 
can hardly be seen as a sign of a divergence (and hence as a signature of an 
in-plane phase transition). Our DFT results seem to completely miss this 
enhanced structural feature in Fig.~\ref{fig:inplane-partial}(c), 
where simulation results clearly show increased structure at larger $q$ values. 
This indicates a compactification of the first layer of ions in agreement with our 
previous out-of-plane analysis. 

%

\begin{figure}
\centering
\includegraphics[width=8.1cm]{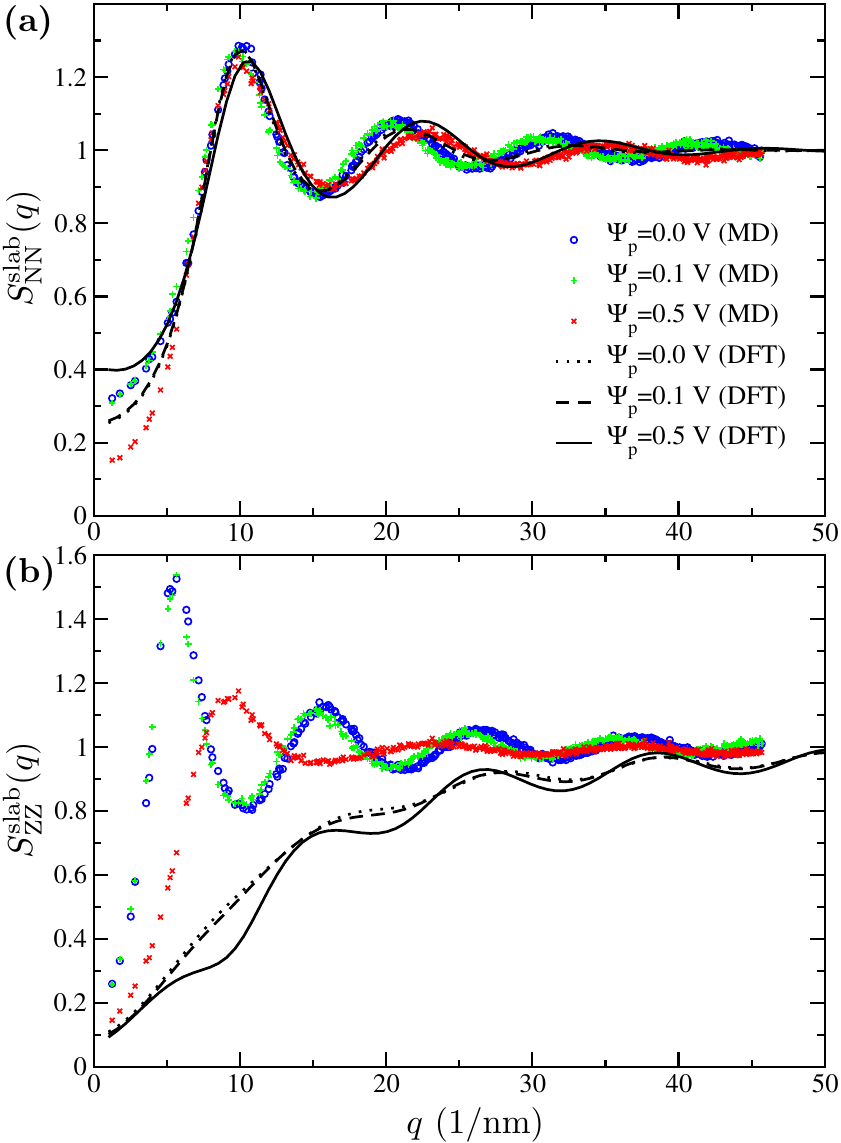}
\caption{\label{fig:inplane-structure-potentials}(Color online) 
The	(a) particle-particle (NN) and (b) charge-charge (ZZ) in-plane structure in the 
first layer of ions at the positive electrode. 
The structure is shown for three electrostatic potentials $\Psi_{\rm p}$ applied to 
the electrode. 
}
\end{figure}


In Figs.~\ref{fig:inplane-structure-potentials}(a) and (b) we show the particle-particle (NN) and 
charge-charge (ZZ) structure, which are linear combinations of the partial in-plane structure factors 
shown in Fig.~\ref{fig:inplane-partial}, respectively. Both are presented 
for the same three potentials as in Fig.~\ref{fig:inplane-partial} and for both DFT and simulations. 
The NN structures are all quite similar (except at low $q$ for the high potential simulations) 
and show a reasonably good agreement between DFT and simulation. This is to be attributed to the high 
quality of the FMT part of the functional which captures overall packing 
effects accurately. In contrast, the ZZ structures show a very poor 
agreement between DFT and simulations, again showing a decay to the large-$q$ 
limit in DFT that is extremely slow compared to the simulations. 

The simulation results in
Fig.~\ref{fig:inplane-structure-potentials}(b) show a significant shift of the 
primary peak from $q\sim6$ nm$^{-1}$ for $\Psi_{\rm p}=0.1$ V to $q\sim10$ 
nm$^{-1}$ for $\Psi_{\rm p}=0.5$ V, consistent with a very strong adsorption of 
ionic charge to the electrode surface. As discussed previously, this finding is 
consistent with the results of Kirchner \textit{et al.} \cite{kirchner_ea110_2013} 
and our findings from Sec.~\ref{sec:quantifying}. 
All in all we argue that the present DFT is not sufficient to study the 
in-plane structure of an ionic liquid in the vicinity of a (highly charged) 
electrode, and further investigations to correct for the slow large-$q$ decay in the 
correlations is required.



%

\section{Concluding discussion}
\label{sec:conclusion}

In this paper we investigated a free-energy functional for a size-asymmetric 
primitive model of spherical ions in order to describe the in- and out-of-plane 
structure of an ionic liquid confined in a parallel-plate capacitor by the means of 
classical DFT. Both the steric repulsions as well as the charges 
are important in these dense and strongly coupled Coulombic systems. The ionic 
hard-core repulsions are accurately treated on the basis of a fundamental 
measure functional whereas the Coulombic interactions are taken into account at 
a mean-field type Poisson-Boltzmann level. To ensure the isolation of problems we 
found for the calculation of in-plane structure from DFT, we neglected further 
(approximate) contributions to the functional, which would correct for correlations 
between the respected hard-sphere and Coulombic contributions, but which are still 
unknown for inhomogeneous systems. 
For several cell voltages we first investigated the ionic density 
profiles at high concentrations and potentials as a function of the distance 
from the electrodes, finding reasonable but far-from-perfect agreement with our 
MD simulations of the same model. The most obvious deviation from our simulation 
results is a much weaker layering of oppositely charged ions in the EDL, observed 
from DFT, while general layering and charge exchange in between layers still is captured. 
Nevertheless, the out-of-plane profiles allow for the calculation of the differential 
charge and molar capacitances, quantities that are relevant for energy-storage and desalination 
devices. An interesting issue concerns the possibility of an in-plane structural 
phase transition associated with anomalies in the differential capacitance as 
observed recently in simulations \cite{merlet_ea101_2013,merlet_jpcc118_2014,rotenberg_jpcl6_2015}. 
While our out-of-plane results show a hint for this structural transition, 
the in-plane structure that follows from the direct correlations of our functional 
gives a very poor account of our simulated in-plane structure. As a main problem, 
we isolated the slow decay of correlations and connected it with the previously 
described underestimation of charge layering. However, our work points towards 
the construction of better functionals, which have to be considered to comprehensively study 
the observed phenomena in ionic liquids confined in (porous) electrodes. A first 
candidate is possibly the MSA-corrected functional that was recently used in 
\cite{mier-y-teran_jcp92_1990,haertel_jpcm27_2015}. 
Although we took the asymmetry between cations and 
anions into account by assigning them different hard-sphere diameters, it is 
also conceivable that the non-spherical shape of some of the ions is a crucial 
ingredient that needs to be incorporated at the level of the hard-core 
functional. After all, the rod-like character of some of the ions in real ionic 
liquids allows for voltage-induced structural changes of the double layer that 
involve alignment of the ions. Given the practical importance of the systems at 
hand and the complexity of the required functionals, classical DFT is facing a 
challenging, interesting, and hopefully bright future to address confined ionic 
liquids in external potentials.

\section*{Acknowledgements}

We thank R. Evans for stimulating discussions. 
This work is part of the D-ITP consortium, a program of the 
Netherlands Organisation for Scientific Research (NWO) that is funded by 
the Dutch Ministry of Education, Culture and Science (OCW). 
We acknowledge financial support by the DFG within priority program SPP 1726 
(grant number SP 1382/3-1), from an NWO-VICI grant, and from the European Union under the Marie 
Sk\l{}odowska-Curie grant agreement No. 656327.


\begin{thebibliography}{73}%
\makeatletter
\providecommand \@ifxundefined [1]{%
 \@ifx{#1\undefined}
}%
\providecommand \@ifnum [1]{%
 \ifnum #1\expandafter \@firstoftwo
 \else \expandafter \@secondoftwo
 \fi
}%
\providecommand \@ifx [1]{%
 \ifx #1\expandafter \@firstoftwo
 \else \expandafter \@secondoftwo
 \fi
}%
\providecommand \natexlab [1]{#1}%
\providecommand \enquote  [1]{``#1''}%
\providecommand \bibnamefont  [1]{#1}%
\providecommand \bibfnamefont [1]{#1}%
\providecommand \citenamefont [1]{#1}%
\providecommand \href@noop [0]{\@secondoftwo}%
\providecommand \href [0]{\begingroup \@sanitize@url \@href}%
\providecommand \@href[1]{\@@startlink{#1}\@@href}%
\providecommand \@@href[1]{\endgroup#1\@@endlink}%
\providecommand \@sanitize@url [0]{\catcode `\\12\catcode `\$12\catcode
  `\&12\catcode `\#12\catcode `\^12\catcode `\_12\catcode `\%12\relax}%
\providecommand \@@startlink[1]{}%
\providecommand \@@endlink[0]{}%
\providecommand \url  [0]{\begingroup\@sanitize@url \@url }%
\providecommand \@url [1]{\endgroup\@href {#1}{\urlprefix }}%
\providecommand \urlprefix  [0]{URL }%
\providecommand \Eprint [0]{\href }%
\providecommand \doibase [0]{http://dx.doi.org/}%
\providecommand \selectlanguage [0]{\@gobble}%
\providecommand \bibinfo  [0]{\@secondoftwo}%
\providecommand \bibfield  [0]{\@secondoftwo}%
\providecommand \translation [1]{[#1]}%
\providecommand \BibitemOpen [0]{}%
\providecommand \bibitemStop [0]{}%
\providecommand \bibitemNoStop [0]{.\EOS\space}%
\providecommand \EOS [0]{\spacefactor3000\relax}%
\providecommand \BibitemShut  [1]{\csname bibitem#1\endcsname}%
\let\auto@bib@innerbib\@empty
\bibitem [{\citenamefont {Simon}\ and\ \citenamefont
  {Gogotsi}(2008)}]{simon_nm7_2008}%
  \BibitemOpen
  \bibfield  {author} {\bibinfo {author} {\bibfnamefont {P.}~\bibnamefont
  {Simon}}\ and\ \bibinfo {author} {\bibfnamefont {Y.}~\bibnamefont
  {Gogotsi}},\ }\href {\doibase 10.1038/nmat2297} {\bibfield  {journal}
  {\bibinfo  {journal} {Nat. Mater.}\ }\textbf {\bibinfo {volume} {7}},\
  \bibinfo {pages} {845} (\bibinfo {year} {2008})}\BibitemShut {NoStop}%
\bibitem [{\citenamefont {Vatamanu}\ and\ \citenamefont
  {Bedrov}(2015)}]{vatamanu_jpcl6_2015}%
  \BibitemOpen
  \bibfield  {author} {\bibinfo {author} {\bibfnamefont {J.}~\bibnamefont
  {Vatamanu}}\ and\ \bibinfo {author} {\bibfnamefont {D.}~\bibnamefont
  {Bedrov}},\ }\href {\doibase 10.1021/acs.jpclett.5b01199} {\bibfield
  {journal} {\bibinfo  {journal} {J. Phys. Chem. Lett.}\ }\textbf {\bibinfo
  {volume} {6}},\ \bibinfo {pages} {3594} (\bibinfo {year} {2015})}\BibitemShut
  {NoStop}%
\bibitem [{\citenamefont {Zhu}\ \emph {et~al.}(2011)\citenamefont {Zhu},
  \citenamefont {Murali}, \citenamefont {Stoller}, \citenamefont {Ganesh},
  \citenamefont {Cai}, \citenamefont {Ferreira}, \citenamefont {Pirkle},
  \citenamefont {Wallace}, \citenamefont {Cychosz}, \citenamefont {Thommes},
  \citenamefont {Su}, \citenamefont {Stach},\ and\ \citenamefont
  {Ruoff}}]{zhu_science332_2011}%
  \BibitemOpen
  \bibfield  {author} {\bibinfo {author} {\bibfnamefont {Y.}~\bibnamefont
  {Zhu}}, \bibinfo {author} {\bibfnamefont {S.}~\bibnamefont {Murali}},
  \bibinfo {author} {\bibfnamefont {M.~D.}\ \bibnamefont {Stoller}}, \bibinfo
  {author} {\bibfnamefont {K.~J.}\ \bibnamefont {Ganesh}}, \bibinfo {author}
  {\bibfnamefont {W.}~\bibnamefont {Cai}}, \bibinfo {author} {\bibfnamefont
  {P.~J.}\ \bibnamefont {Ferreira}}, \bibinfo {author} {\bibfnamefont
  {A.}~\bibnamefont {Pirkle}}, \bibinfo {author} {\bibfnamefont {R.~M.}\
  \bibnamefont {Wallace}}, \bibinfo {author} {\bibfnamefont {K.~A.}\
  \bibnamefont {Cychosz}}, \bibinfo {author} {\bibfnamefont {M.}~\bibnamefont
  {Thommes}}, \bibinfo {author} {\bibfnamefont {D.}~\bibnamefont {Su}},
  \bibinfo {author} {\bibfnamefont {E.~A.}\ \bibnamefont {Stach}}, \ and\
  \bibinfo {author} {\bibfnamefont {R.~S.}\ \bibnamefont {Ruoff}},\ }\href@noop
  {} {\bibfield  {journal} {\bibinfo  {journal} {Science}\ }\textbf {\bibinfo
  {volume} {332}},\ \bibinfo {pages} {1537} (\bibinfo {year}
  {2011})}\BibitemShut {NoStop}%
\bibitem [{\citenamefont {Brogioli}(2009)}]{brogioli_prl103_2009}%
  \BibitemOpen
  \bibfield  {author} {\bibinfo {author} {\bibfnamefont {D.}~\bibnamefont
  {Brogioli}},\ }\href {\doibase 10.1103/PhysRevLett.103.058501} {\bibfield
  {journal} {\bibinfo  {journal} {Phys. Rev. Lett.}\ }\textbf {\bibinfo
  {volume} {103}},\ \bibinfo {pages} {058501} (\bibinfo {year}
  {2009})}\BibitemShut {NoStop}%
\bibitem [{\citenamefont {Boon}\ and\ \citenamefont {van
  Roij}(2011)}]{boon_mp109_2011}%
  \BibitemOpen
  \bibfield  {author} {\bibinfo {author} {\bibfnamefont {N.}~\bibnamefont
  {Boon}}\ and\ \bibinfo {author} {\bibfnamefont {R.}~\bibnamefont {van
  Roij}},\ }\href {\doibase 10.1080/00268976.2011.554334} {\bibfield  {journal}
  {\bibinfo  {journal} {Mol. Phys.}\ }\textbf {\bibinfo {volume} {109}},\
  \bibinfo {pages} {1229} (\bibinfo {year} {2011})}\BibitemShut {NoStop}%
\bibitem [{\citenamefont {Rica}\ \emph {et~al.}(2013)\citenamefont {Rica},
  \citenamefont {Ziano}, \citenamefont {Salerno}, \citenamefont {Mantegazza},
  \citenamefont {van Roij},\ and\ \citenamefont
  {Brogioli}}]{rica_entropy15_2013}%
  \BibitemOpen
  \bibfield  {author} {\bibinfo {author} {\bibfnamefont {R.~A.}\ \bibnamefont
  {Rica}}, \bibinfo {author} {\bibfnamefont {R.}~\bibnamefont {Ziano}},
  \bibinfo {author} {\bibfnamefont {D.}~\bibnamefont {Salerno}}, \bibinfo
  {author} {\bibfnamefont {F.}~\bibnamefont {Mantegazza}}, \bibinfo {author}
  {\bibfnamefont {R.}~\bibnamefont {van Roij}}, \ and\ \bibinfo {author}
  {\bibfnamefont {D.}~\bibnamefont {Brogioli}},\ }\href {\doibase
  10.3390/e15041388} {\bibfield  {journal} {\bibinfo  {journal} {Entropy}\
  }\textbf {\bibinfo {volume} {15}},\ \bibinfo {pages} {1388} (\bibinfo {year}
  {2013})}\BibitemShut {NoStop}%
\bibitem [{\citenamefont {Hamelers}\ \emph {et~al.}(2014)\citenamefont
  {Hamelers}, \citenamefont {Schaetzle}, \citenamefont {Paz-Garc\'ia},
  \citenamefont {Biesheuvel},\ and\ \citenamefont
  {Buisman}}]{hamelers_estl1_2014}%
  \BibitemOpen
  \bibfield  {author} {\bibinfo {author} {\bibfnamefont {H.~V.~M.}\
  \bibnamefont {Hamelers}}, \bibinfo {author} {\bibfnamefont {O.}~\bibnamefont
  {Schaetzle}}, \bibinfo {author} {\bibfnamefont {J.~M.}\ \bibnamefont
  {Paz-Garc\'ia}}, \bibinfo {author} {\bibfnamefont {P.~M.}\ \bibnamefont
  {Biesheuvel}}, \ and\ \bibinfo {author} {\bibfnamefont {C.~J.~N.}\
  \bibnamefont {Buisman}},\ }\href {\doibase 10.1021/ez4000059} {\bibfield
  {journal} {\bibinfo  {journal} {Environ. Sci. Technol. Lett.}\ }\textbf
  {\bibinfo {volume} {1}},\ \bibinfo {pages} {31} (\bibinfo {year}
  {2014})}\BibitemShut {NoStop}%
\bibitem [{\citenamefont {H\"artel}\ \emph
  {et~al.}(2015{\natexlab{a}})\citenamefont {H\"artel}, \citenamefont
  {Janssen}, \citenamefont {Weingarth}, \citenamefont {Presser},\ and\
  \citenamefont {van Roij}}]{haertel_ees8_2015}%
  \BibitemOpen
  \bibfield  {author} {\bibinfo {author} {\bibfnamefont {A.}~\bibnamefont
  {H\"artel}}, \bibinfo {author} {\bibfnamefont {M.}~\bibnamefont {Janssen}},
  \bibinfo {author} {\bibfnamefont {D.}~\bibnamefont {Weingarth}}, \bibinfo
  {author} {\bibfnamefont {V.}~\bibnamefont {Presser}}, \ and\ \bibinfo
  {author} {\bibfnamefont {R.}~\bibnamefont {van Roij}},\ }\href {\doibase
  10.1039/C5EE01192B} {\bibfield  {journal} {\bibinfo  {journal} {Energy
  Environ. Sci.}\ }\textbf {\bibinfo {volume} {8}},\ \bibinfo {pages} {2396}
  (\bibinfo {year} {2015}{\natexlab{a}})}\BibitemShut {NoStop}%
\bibitem [{\citenamefont {Porada}\ \emph {et~al.}(2013)\citenamefont {Porada},
  \citenamefont {Borchardt}, \citenamefont {Oschatz}, \citenamefont {Bryjak},
  \citenamefont {Atchison}, \citenamefont {Keesman}, \citenamefont {Kaskel},
  \citenamefont {Biesheuvel},\ and\ \citenamefont
  {Presser}}]{porada_ees6_2013}%
  \BibitemOpen
  \bibfield  {author} {\bibinfo {author} {\bibfnamefont {S.}~\bibnamefont
  {Porada}}, \bibinfo {author} {\bibfnamefont {L.}~\bibnamefont {Borchardt}},
  \bibinfo {author} {\bibfnamefont {M.}~\bibnamefont {Oschatz}}, \bibinfo
  {author} {\bibfnamefont {M.}~\bibnamefont {Bryjak}}, \bibinfo {author}
  {\bibfnamefont {J.~S.}\ \bibnamefont {Atchison}}, \bibinfo {author}
  {\bibfnamefont {K.~J.}\ \bibnamefont {Keesman}}, \bibinfo {author}
  {\bibfnamefont {S.}~\bibnamefont {Kaskel}}, \bibinfo {author} {\bibfnamefont
  {P.~M.}\ \bibnamefont {Biesheuvel}}, \ and\ \bibinfo {author} {\bibfnamefont
  {V.}~\bibnamefont {Presser}},\ }\href {\doibase 10.1039/c3ee42209g}
  {\bibfield  {journal} {\bibinfo  {journal} {Energy Environ. Sci.}\ }\textbf
  {\bibinfo {volume} {6}},\ \bibinfo {pages} {3700} (\bibinfo {year}
  {2013})}\BibitemShut {NoStop}%
\bibitem [{\citenamefont {Suss}\ \emph {et~al.}(2015)\citenamefont {Suss},
  \citenamefont {Porada}, \citenamefont {Sun}, \citenamefont {Biesheuvel},
  \citenamefont {Yoon},\ and\ \citenamefont {Presser}}]{suss_ees8_2015}%
  \BibitemOpen
  \bibfield  {author} {\bibinfo {author} {\bibfnamefont {M.~E.}\ \bibnamefont
  {Suss}}, \bibinfo {author} {\bibfnamefont {S.}~\bibnamefont {Porada}},
  \bibinfo {author} {\bibfnamefont {X.}~\bibnamefont {Sun}}, \bibinfo {author}
  {\bibfnamefont {P.~M.}\ \bibnamefont {Biesheuvel}}, \bibinfo {author}
  {\bibfnamefont {J.}~\bibnamefont {Yoon}}, \ and\ \bibinfo {author}
  {\bibfnamefont {V.}~\bibnamefont {Presser}},\ }\href {\doibase
  10.1039/C5EE00519A} {\bibfield  {journal} {\bibinfo  {journal} {Energy
  Environ. Sci.}\ }\textbf {\bibinfo {volume} {8}},\ \bibinfo {pages} {2296}
  (\bibinfo {year} {2015})}\BibitemShut {NoStop}%
\bibitem [{\citenamefont {Rogers}\ and\ \citenamefont
  {Seddon}(2003)}]{rogers_science302_2003}%
  \BibitemOpen
  \bibfield  {author} {\bibinfo {author} {\bibfnamefont {R.~D.}\ \bibnamefont
  {Rogers}}\ and\ \bibinfo {author} {\bibfnamefont {K.~R.}\ \bibnamefont
  {Seddon}},\ }\href {\doibase 10.1126/science.1090313} {\bibfield  {journal}
  {\bibinfo  {journal} {Science}\ }\textbf {\bibinfo {volume} {302}},\ \bibinfo
  {pages} {792} (\bibinfo {year} {2003})}\BibitemShut {NoStop}%
\bibitem [{\citenamefont {Oleksy}\ and\ \citenamefont
  {Hansen}(2006)}]{oleksy_mp104_2006}%
  \BibitemOpen
  \bibfield  {author} {\bibinfo {author} {\bibfnamefont {A.}~\bibnamefont
  {Oleksy}}\ and\ \bibinfo {author} {\bibfnamefont {J.-P.}\ \bibnamefont
  {Hansen}},\ }\href {\doibase 10.1080/00268970600864491} {\bibfield  {journal}
  {\bibinfo  {journal} {Mol. Phys.}\ }\textbf {\bibinfo {volume} {104}},\
  \bibinfo {pages} {2871} (\bibinfo {year} {2006})}\BibitemShut {NoStop}%
\bibitem [{\citenamefont {Kornyshev}(2007)}]{kornyshev2007double}%
  \BibitemOpen
  \bibfield  {author} {\bibinfo {author} {\bibfnamefont {A.~A.}\ \bibnamefont
  {Kornyshev}},\ }\href@noop {} {\bibfield  {journal} {\bibinfo  {journal} {J.
  Phys. Chem. B}\ }\textbf {\bibinfo {volume} {111}},\ \bibinfo {pages} {5545}
  (\bibinfo {year} {2007})}\BibitemShut {NoStop}%
\bibitem [{\citenamefont {Fedorov}\ and\ \citenamefont
  {Kornyshev}(2008)}]{fedorov_jpcb112_2008}%
  \BibitemOpen
  \bibfield  {author} {\bibinfo {author} {\bibfnamefont {M.~V.}\ \bibnamefont
  {Fedorov}}\ and\ \bibinfo {author} {\bibfnamefont {A.~A.}\ \bibnamefont
  {Kornyshev}},\ }\href {\doibase 10.1021/jp803440q} {\bibfield  {journal}
  {\bibinfo  {journal} {J. Phys. Chem. B}\ }\textbf {\bibinfo {volume} {112}},\
  \bibinfo {pages} {11868} (\bibinfo {year} {2008})}\BibitemShut {NoStop}%
\bibitem [{\citenamefont {Georgi}\ \emph {et~al.}(2010)\citenamefont {Georgi},
  \citenamefont {Kornyshev},\ and\ \citenamefont
  {Fedorov}}]{georgi_jec649_2010}%
  \BibitemOpen
  \bibfield  {author} {\bibinfo {author} {\bibfnamefont {N.}~\bibnamefont
  {Georgi}}, \bibinfo {author} {\bibfnamefont {A.}~\bibnamefont {Kornyshev}}, \
  and\ \bibinfo {author} {\bibfnamefont {M.}~\bibnamefont {Fedorov}},\ }\href
  {\doibase 10.1016/j.jelechem.2010.07.004} {\bibfield  {journal} {\bibinfo
  {journal} {J. Electroanal. Chem.}\ }\textbf {\bibinfo {volume} {649}},\
  \bibinfo {pages} {261} (\bibinfo {year} {2010})}\BibitemShut {NoStop}%
\bibitem [{\citenamefont {Jiang}\ \emph {et~al.}(2011)\citenamefont {Jiang},
  \citenamefont {Meng},\ and\ \citenamefont {Wu}}]{jiang_cpl504_2011}%
  \BibitemOpen
  \bibfield  {author} {\bibinfo {author} {\bibfnamefont {D.}~\bibnamefont
  {Jiang}}, \bibinfo {author} {\bibfnamefont {D.}~\bibnamefont {Meng}}, \ and\
  \bibinfo {author} {\bibfnamefont {J.}~\bibnamefont {Wu}},\ }\href {\doibase
  10.1016/j.cplett.2011.01.072} {\bibfield  {journal} {\bibinfo  {journal}
  {Chem. Phys. Lett.}\ }\textbf {\bibinfo {volume} {504}},\ \bibinfo {pages}
  {153} (\bibinfo {year} {2011})}\BibitemShut {NoStop}%
\bibitem [{\citenamefont {Henderson}\ \emph {et~al.}(2012)\citenamefont
  {Henderson}, \citenamefont {Jiang}, \citenamefont {Jin},\ and\ \citenamefont
  {Wu}}]{henderson_jpcb116_2012}%
  \BibitemOpen
  \bibfield  {author} {\bibinfo {author} {\bibfnamefont {D.}~\bibnamefont
  {Henderson}}, \bibinfo {author} {\bibfnamefont {D.}~\bibnamefont {Jiang}},
  \bibinfo {author} {\bibfnamefont {Z.}~\bibnamefont {Jin}}, \ and\ \bibinfo
  {author} {\bibfnamefont {J.}~\bibnamefont {Wu}},\ }\href {\doibase
  10.1021/jp305400z} {\bibfield  {journal} {\bibinfo  {journal} {J. Phys. Chem.
  B}\ }\textbf {\bibinfo {volume} {116}},\ \bibinfo {pages} {11356} (\bibinfo
  {year} {2012})}\BibitemShut {NoStop}%
\bibitem [{\citenamefont {Lamperski}\ \emph {et~al.}(2013)\citenamefont
  {Lamperski}, \citenamefont {Kaja}, \citenamefont {Bhuiyan}, \citenamefont
  {Wu},\ and\ \citenamefont {Henderson}}]{lamperski_jcp139_2013}%
  \BibitemOpen
  \bibfield  {author} {\bibinfo {author} {\bibfnamefont {S.}~\bibnamefont
  {Lamperski}}, \bibinfo {author} {\bibfnamefont {M.}~\bibnamefont {Kaja}},
  \bibinfo {author} {\bibfnamefont {L.~B.}\ \bibnamefont {Bhuiyan}}, \bibinfo
  {author} {\bibfnamefont {J.}~\bibnamefont {Wu}}, \ and\ \bibinfo {author}
  {\bibfnamefont {D.}~\bibnamefont {Henderson}},\ }\href {\doibase
  http://dx.doi.org/10.1063/1.4817325} {\bibfield  {journal} {\bibinfo
  {journal} {J. Chem. Phys.}\ }\textbf {\bibinfo {volume} {139}},\ \bibinfo
  {pages} {054703} (\bibinfo {year} {2013})}\BibitemShut {NoStop}%
\bibitem [{\citenamefont {Breitsprecher}\ \emph {et~al.}(2014)\citenamefont
  {Breitsprecher}, \citenamefont {Ko\u{s}ovan},\ and\ \citenamefont
  {Holm}}]{breitsprecher_jpcm26_2014}%
  \BibitemOpen
  \bibfield  {author} {\bibinfo {author} {\bibfnamefont {K.}~\bibnamefont
  {Breitsprecher}}, \bibinfo {author} {\bibfnamefont {P.}~\bibnamefont
  {Ko\u{s}ovan}}, \ and\ \bibinfo {author} {\bibfnamefont {C.}~\bibnamefont
  {Holm}},\ }\href {\doibase 10.1088/0953-8984/26/28/284108} {\bibfield
  {journal} {\bibinfo  {journal} {J. Phys.: Condens. Matter}\ }\textbf
  {\bibinfo {volume} {26}},\ \bibinfo {pages} {284108} (\bibinfo {year}
  {2014})}\BibitemShut {NoStop}%
\bibitem [{\citenamefont {Han}\ \emph {et~al.}(2014)\citenamefont {Han},
  \citenamefont {Huang},\ and\ \citenamefont {Yan}}]{han_jpcm26_2014}%
  \BibitemOpen
  \bibfield  {author} {\bibinfo {author} {\bibfnamefont {Y.}~\bibnamefont
  {Han}}, \bibinfo {author} {\bibfnamefont {S.}~\bibnamefont {Huang}}, \ and\
  \bibinfo {author} {\bibfnamefont {T.}~\bibnamefont {Yan}},\ }\href {\doibase
  10.1088/0953-8984/26/28/284103} {\bibfield  {journal} {\bibinfo  {journal}
  {J. Phys.: Condens. Matter}\ }\textbf {\bibinfo {volume} {26}},\ \bibinfo
  {pages} {284103} (\bibinfo {year} {2014})}\BibitemShut {NoStop}%
\bibitem [{\citenamefont {Kong}\ \emph {et~al.}(2015)\citenamefont {Kong},
  \citenamefont {Wu},\ and\ \citenamefont {Henderson}}]{kong_jcis449_2015}%
  \BibitemOpen
  \bibfield  {author} {\bibinfo {author} {\bibfnamefont {X.}~\bibnamefont
  {Kong}}, \bibinfo {author} {\bibfnamefont {J.}~\bibnamefont {Wu}}, \ and\
  \bibinfo {author} {\bibfnamefont {D.}~\bibnamefont {Henderson}},\ }\href
  {\doibase http://dx.doi.org/10.1016/j.jcis.2014.11.012} {\bibfield  {journal}
  {\bibinfo  {journal} {J. Colloid Interface Sci.}\ }\textbf {\bibinfo {volume}
  {449}},\ \bibinfo {pages} {130} (\bibinfo {year} {2015})}\BibitemShut
  {NoStop}%
\bibitem [{\citenamefont {Merlet}\ \emph {et~al.}(2013)\citenamefont {Merlet},
  \citenamefont {Salanne}, \citenamefont {Rotenberg},\ and\ \citenamefont
  {Madden}}]{merlet_ea101_2013}%
  \BibitemOpen
  \bibfield  {author} {\bibinfo {author} {\bibfnamefont {C.}~\bibnamefont
  {Merlet}}, \bibinfo {author} {\bibfnamefont {M.}~\bibnamefont {Salanne}},
  \bibinfo {author} {\bibfnamefont {B.}~\bibnamefont {Rotenberg}}, \ and\
  \bibinfo {author} {\bibfnamefont {P.~A.}\ \bibnamefont {Madden}},\ }\href
  {\doibase http://dx.doi.org/10.1016/j.electacta.2012.12.107} {\bibfield
  {journal} {\bibinfo  {journal} {Electrochim. Acta}\ }\textbf {\bibinfo
  {volume} {101}},\ \bibinfo {pages} {262} (\bibinfo {year}
  {2013})}\BibitemShut {NoStop}%
\bibitem [{\citenamefont {Merlet}\ \emph {et~al.}(2014)\citenamefont {Merlet},
  \citenamefont {Limmer}, \citenamefont {Salanne}, \citenamefont {van Roij},
  \citenamefont {Madden}, \citenamefont {Chandler},\ and\ \citenamefont
  {Rotenberg}}]{merlet_jpcc118_2014}%
  \BibitemOpen
  \bibfield  {author} {\bibinfo {author} {\bibfnamefont {C.}~\bibnamefont
  {Merlet}}, \bibinfo {author} {\bibfnamefont {D.~T.}\ \bibnamefont {Limmer}},
  \bibinfo {author} {\bibfnamefont {M.}~\bibnamefont {Salanne}}, \bibinfo
  {author} {\bibfnamefont {R.}~\bibnamefont {van Roij}}, \bibinfo {author}
  {\bibfnamefont {P.~A.}\ \bibnamefont {Madden}}, \bibinfo {author}
  {\bibfnamefont {D.}~\bibnamefont {Chandler}}, \ and\ \bibinfo {author}
  {\bibfnamefont {B.}~\bibnamefont {Rotenberg}},\ }\href {\doibase
  10.1021/jp503224w} {\bibfield  {journal} {\bibinfo  {journal} {J. Phys. Chem.
  C}\ }\textbf {\bibinfo {volume} {118}},\ \bibinfo {pages} {18291} (\bibinfo
  {year} {2014})}\BibitemShut {NoStop}%
\bibitem [{\citenamefont {Kirchner}\ \emph {et~al.}(2013)\citenamefont
  {Kirchner}, \citenamefont {Kirchner}, \citenamefont {Ivani\u{s}t\u{e}ev},\
  and\ \citenamefont {Fedorov}}]{kirchner_ea110_2013}%
  \BibitemOpen
  \bibfield  {author} {\bibinfo {author} {\bibfnamefont {K.}~\bibnamefont
  {Kirchner}}, \bibinfo {author} {\bibfnamefont {T.}~\bibnamefont {Kirchner}},
  \bibinfo {author} {\bibfnamefont {V.}~\bibnamefont {Ivani\u{s}t\u{e}ev}}, \
  and\ \bibinfo {author} {\bibfnamefont {M.}~\bibnamefont {Fedorov}},\ }\href
  {\doibase http://dx.doi.org/10.1016/j.electacta.2013.05.049} {\bibfield
  {journal} {\bibinfo  {journal} {Electrochim. Acta}\ }\textbf {\bibinfo
  {volume} {110}},\ \bibinfo {pages} {762} (\bibinfo {year}
  {2013})}\BibitemShut {NoStop}%
\bibitem [{\citenamefont {Rotenberg}\ and\ \citenamefont
  {Salanne}(2015)}]{rotenberg_jpcl6_2015}%
  \BibitemOpen
  \bibfield  {author} {\bibinfo {author} {\bibfnamefont {B.}~\bibnamefont
  {Rotenberg}}\ and\ \bibinfo {author} {\bibfnamefont {M.}~\bibnamefont
  {Salanne}},\ }\href {\doibase 10.1021/acs.jpclett.5b01889} {\bibfield
  {journal} {\bibinfo  {journal} {J. Phys. Chem. Lett.}\ }\textbf {\bibinfo
  {volume} {6}},\ \bibinfo {pages} {4978} (\bibinfo {year} {2015})}\BibitemShut
  {NoStop}%
\bibitem [{\citenamefont {Jeon}\ \emph {et~al.}(2012)\citenamefont {Jeon},
  \citenamefont {Vaknin}, \citenamefont {Bu}, \citenamefont {Sung},
  \citenamefont {Ouchi}, \citenamefont {Sung},\ and\ \citenamefont
  {Kim}}]{jeon_prl108_2012}%
  \BibitemOpen
  \bibfield  {author} {\bibinfo {author} {\bibfnamefont {Y.}~\bibnamefont
  {Jeon}}, \bibinfo {author} {\bibfnamefont {D.}~\bibnamefont {Vaknin}},
  \bibinfo {author} {\bibfnamefont {W.}~\bibnamefont {Bu}}, \bibinfo {author}
  {\bibfnamefont {J.}~\bibnamefont {Sung}}, \bibinfo {author} {\bibfnamefont
  {Y.}~\bibnamefont {Ouchi}}, \bibinfo {author} {\bibfnamefont
  {W.}~\bibnamefont {Sung}}, \ and\ \bibinfo {author} {\bibfnamefont
  {D.}~\bibnamefont {Kim}},\ }\href {\doibase 10.1103/PhysRevLett.108.055502}
  {\bibfield  {journal} {\bibinfo  {journal} {Phys. Rev. Lett.}\ }\textbf
  {\bibinfo {volume} {108}},\ \bibinfo {pages} {055502} (\bibinfo {year}
  {2012})}\BibitemShut {NoStop}%
\bibitem [{\citenamefont {Griffin}\ \emph {et~al.}(2015)\citenamefont
  {Griffin}, \citenamefont {Forse}, \citenamefont {Tsai}, \citenamefont
  {Taberna}, \citenamefont {Simon},\ and\ \citenamefont
  {Grey}}]{griffin_nm14_2015}%
  \BibitemOpen
  \bibfield  {author} {\bibinfo {author} {\bibfnamefont {J.~M.}\ \bibnamefont
  {Griffin}}, \bibinfo {author} {\bibfnamefont {A.~C.}\ \bibnamefont {Forse}},
  \bibinfo {author} {\bibfnamefont {W.-Y.}\ \bibnamefont {Tsai}}, \bibinfo
  {author} {\bibfnamefont {P.-L.}\ \bibnamefont {Taberna}}, \bibinfo {author}
  {\bibfnamefont {P.}~\bibnamefont {Simon}}, \ and\ \bibinfo {author}
  {\bibfnamefont {C.~P.}\ \bibnamefont {Grey}},\ }\href {\doibase
  10.1038/nmat4318} {\bibfield  {journal} {\bibinfo  {journal} {Nat. Mater.}\
  }\textbf {\bibinfo {volume} {14}},\ \bibinfo {pages} {812} (\bibinfo {year}
  {2015})}\BibitemShut {NoStop}%
\bibitem [{\citenamefont {Baus}(1983)}]{baus_molphys48_1983}%
  \BibitemOpen
  \bibfield  {author} {\bibinfo {author} {\bibfnamefont {M.}~\bibnamefont
  {Baus}},\ }\href {\doibase 10.1080/00268978300100261} {\bibfield  {journal}
  {\bibinfo  {journal} {Mol. Phys.}\ }\textbf {\bibinfo {volume} {48}},\
  \bibinfo {pages} {347} (\bibinfo {year} {1983})}\BibitemShut {NoStop}%
\bibitem [{\citenamefont {Hohenberg}\ and\ \citenamefont
  {Kohn}(1964)}]{hohenberg_pr136_1964}%
  \BibitemOpen
  \bibfield  {author} {\bibinfo {author} {\bibfnamefont {P.}~\bibnamefont
  {Hohenberg}}\ and\ \bibinfo {author} {\bibfnamefont {W.}~\bibnamefont
  {Kohn}},\ }\href {\doibase 10.1103/PhysRev.136.B864} {\bibfield  {journal}
  {\bibinfo  {journal} {Phys. Rev.}\ }\textbf {\bibinfo {volume} {136}},\
  \bibinfo {pages} {B864} (\bibinfo {year} {1964})}\BibitemShut {NoStop}%
\bibitem [{\citenamefont {Jones}(2015)}]{jones_rmp87_2015}%
  \BibitemOpen
  \bibfield  {author} {\bibinfo {author} {\bibfnamefont {R.~O.}\ \bibnamefont
  {Jones}},\ }\href {\doibase 10.1103/RevModPhys.87.897} {\bibfield  {journal}
  {\bibinfo  {journal} {Rev. Mod. Phys.}\ }\textbf {\bibinfo {volume} {87}},\
  \bibinfo {pages} {897} (\bibinfo {year} {2015})}\BibitemShut {NoStop}%
\bibitem [{\citenamefont {Mermin}(1965)}]{mermin_pr137_1965}%
  \BibitemOpen
  \bibfield  {author} {\bibinfo {author} {\bibfnamefont {N.~D.}\ \bibnamefont
  {Mermin}},\ }\href {\doibase 10.1103/PhysRev.137.A1441} {\bibfield  {journal}
  {\bibinfo  {journal} {Phys. Rev.}\ }\textbf {\bibinfo {volume} {137}},\
  \bibinfo {pages} {A1441} (\bibinfo {year} {1965})}\BibitemShut {NoStop}%
\bibitem [{\citenamefont {Ebner}\ \emph {et~al.}(1976)\citenamefont {Ebner},
  \citenamefont {Saam},\ and\ \citenamefont {Stroud}}]{ebner_pra14_1976}%
  \BibitemOpen
  \bibfield  {author} {\bibinfo {author} {\bibfnamefont {C.}~\bibnamefont
  {Ebner}}, \bibinfo {author} {\bibfnamefont {W.~F.}\ \bibnamefont {Saam}}, \
  and\ \bibinfo {author} {\bibfnamefont {D.}~\bibnamefont {Stroud}},\ }\href
  {\doibase 10.1103/PhysRevA.14.2264} {\bibfield  {journal} {\bibinfo
  {journal} {Phys.\ Rev.\ A}\ }\textbf {\bibinfo {volume} {14}},\ \bibinfo
  {pages} {2264} (\bibinfo {year} {1976})}\BibitemShut {NoStop}%
\bibitem [{\citenamefont {Evans}(1979)}]{evans_ap28_1979}%
  \BibitemOpen
  \bibfield  {author} {\bibinfo {author} {\bibfnamefont {R.}~\bibnamefont
  {Evans}},\ }\href {\doibase 10.1080/00018737900101365} {\bibfield  {journal}
  {\bibinfo  {journal} {Adv.\ Phys.}\ }\textbf {\bibinfo {volume} {28}},\
  \bibinfo {pages} {143} (\bibinfo {year} {1979})}\BibitemShut {NoStop}%
\bibitem [{\citenamefont {Tarazona}\ \emph {et~al.}(2008)\citenamefont
  {Tarazona}, \citenamefont {Cuesta},\ and\ \citenamefont
  {Mart\'inez-Rat\'on}}]{tarazona_inbook_2008}%
  \BibitemOpen
  \bibfield  {author} {\bibinfo {author} {\bibfnamefont {P.}~\bibnamefont
  {Tarazona}}, \bibinfo {author} {\bibfnamefont {J.~A.}\ \bibnamefont
  {Cuesta}}, \ and\ \bibinfo {author} {\bibfnamefont {Y.}~\bibnamefont
  {Mart\'inez-Rat\'on}},\ }\enquote {\bibinfo {title} {Density functional
  theories of hard particle systems},}\ in\ \href {\doibase
  10.1007/978-3-540-78767-9} {\emph {\bibinfo {booktitle} {Theory and
  Simulation of Hard-Sphere Fluids and Related Systems}}},\ \bibinfo {series}
  {Lecture Notes in Physics}, Vol.\ \bibinfo {volume} {753},\ \bibinfo {editor}
  {edited by\ \bibinfo {editor} {\bibfnamefont {A.}~\bibnamefont {Mulero}}}\
  (\bibinfo  {publisher} {Springer-Verlag},\ \bibinfo {address} {Berlin
  Heidelberg},\ \bibinfo {year} {2008})\ Chap.~\bibinfo {chapter} {7}, pp.\
  \bibinfo {pages} {247--341}\BibitemShut {NoStop}%
\bibitem [{\citenamefont {Wu}(2006)}]{wu_aiche52_2006}%
  \BibitemOpen
  \bibfield  {author} {\bibinfo {author} {\bibfnamefont {J.}~\bibnamefont
  {Wu}},\ }\href {\doibase 10.1002/aic.10713} {\bibfield  {journal} {\bibinfo
  {journal} {AIChE J.}\ }\textbf {\bibinfo {volume} {52}},\ \bibinfo {pages}
  {1169} (\bibinfo {year} {2006})}\BibitemShut {NoStop}%
\bibitem [{\citenamefont {Hansen}\ and\ \citenamefont
  {McDonald}(2013)}]{hansen_book_2013}%
  \BibitemOpen
  \bibfield  {author} {\bibinfo {author} {\bibfnamefont {J.-P.}\ \bibnamefont
  {Hansen}}\ and\ \bibinfo {author} {\bibfnamefont {I.~R.}\ \bibnamefont
  {McDonald}},\ }\href@noop {} {\emph {\bibinfo {title} {Theory of simple
  liquids}}},\ \bibinfo {edition} {4th}\ ed.\ (\bibinfo  {publisher}
  {Elsevier},\ \bibinfo {year} {2013})\BibitemShut {NoStop}%
\bibitem [{\citenamefont {Seaton}\ \emph {et~al.}(1989)\citenamefont {Seaton},
  \citenamefont {Walton},\ and\ \citenamefont {Quirke}}]{seaton_carbon27_1989}%
  \BibitemOpen
  \bibfield  {author} {\bibinfo {author} {\bibfnamefont {N.~A.}\ \bibnamefont
  {Seaton}}, \bibinfo {author} {\bibfnamefont {J.~P. R.~B.}\ \bibnamefont
  {Walton}}, \ and\ \bibinfo {author} {\bibfnamefont {N.}~\bibnamefont
  {Quirke}},\ }\href {\doibase 10.1016/0008-6223(89)90035-3} {\bibfield
  {journal} {\bibinfo  {journal} {Carbon}\ }\textbf {\bibinfo {volume} {27}},\
  \bibinfo {pages} {853} (\bibinfo {year} {1989})}\BibitemShut {NoStop}%
\bibitem [{\citenamefont {Thommes}(2004)}]{thommes_inbook_2004}%
  \BibitemOpen
  \bibfield  {author} {\bibinfo {author} {\bibfnamefont {M.}~\bibnamefont
  {Thommes}},\ }\enquote {\bibinfo {title} {Physical adsorption
  characterization of ordered and amorphous mesoporous materials},}\ in\
  \href@noop {} {\emph {\bibinfo {booktitle} {Nanoporous Materials: Science and
  Engineering}}},\ \bibinfo {series} {Series on Chemical Engineering},
  Vol.~\bibinfo {volume} {4},\ \bibinfo {editor} {edited by\ \bibinfo {editor}
  {\bibfnamefont {G.~Q.}\ \bibnamefont {Lu}}\ and\ \bibinfo {editor}
  {\bibfnamefont {X.~S.}\ \bibnamefont {Zhao}}}\ (\bibinfo  {publisher}
  {Imperial College Press},\ \bibinfo {address} {London},\ \bibinfo {year}
  {2004})\ Chap.~\bibinfo {chapter} {11}, p.\ \bibinfo {pages}
  {317}\BibitemShut {NoStop}%
\bibitem [{\citenamefont {Ravikovitch}\ and\ \citenamefont
  {Neimark}(2006)}]{ravikovitch_langmuir22_2006}%
  \BibitemOpen
  \bibfield  {author} {\bibinfo {author} {\bibfnamefont {P.~I.}\ \bibnamefont
  {Ravikovitch}}\ and\ \bibinfo {author} {\bibfnamefont {A.~V.}\ \bibnamefont
  {Neimark}},\ }\href {\doibase 10.1021/la0616146} {\bibfield  {journal}
  {\bibinfo  {journal} {Langmuir}\ }\textbf {\bibinfo {volume} {22}},\ \bibinfo
  {pages} {11171} (\bibinfo {year} {2006})}\BibitemShut {NoStop}%
\bibitem [{\citenamefont {Rosenfeld}(1989)}]{rosenfeld_prl63_1989}%
  \BibitemOpen
  \bibfield  {author} {\bibinfo {author} {\bibfnamefont {Y.}~\bibnamefont
  {Rosenfeld}},\ }\href {\doibase 10.1103/PhysRevLett.63.980} {\bibfield
  {journal} {\bibinfo  {journal} {Phys.\ Rev.\ Lett.}\ }\textbf {\bibinfo
  {volume} {63}},\ \bibinfo {pages} {980} (\bibinfo {year} {1989})}\BibitemShut
  {NoStop}%
\bibitem [{\citenamefont {Roth}(2010)}]{roth_jpcm22_2010}%
  \BibitemOpen
  \bibfield  {author} {\bibinfo {author} {\bibfnamefont {R.}~\bibnamefont
  {Roth}},\ }\href {\doibase 10.1088/0953-8984/22/6/063102} {\bibfield
  {journal} {\bibinfo  {journal} {J.\ Phys.: Condens.\ Matter}\ }\textbf
  {\bibinfo {volume} {22}},\ \bibinfo {pages} {063102} (\bibinfo {year}
  {2010})}\BibitemShut {NoStop}%
\bibitem [{\citenamefont {Marechal}\ \emph {et~al.}(2014)\citenamefont
  {Marechal}, \citenamefont {Korden},\ and\ \citenamefont
  {Mecke}}]{marechal_pre90_2014}%
  \BibitemOpen
  \bibfield  {author} {\bibinfo {author} {\bibfnamefont {M.}~\bibnamefont
  {Marechal}}, \bibinfo {author} {\bibfnamefont {S.}~\bibnamefont {Korden}}, \
  and\ \bibinfo {author} {\bibfnamefont {K.}~\bibnamefont {Mecke}},\ }\href
  {\doibase 10.1103/PhysRevE.90.042131} {\bibfield  {journal} {\bibinfo
  {journal} {Phys. Rev. E}\ }\textbf {\bibinfo {volume} {90}},\ \bibinfo
  {pages} {042131} (\bibinfo {year} {2014})}\BibitemShut {NoStop}%
\bibitem [{\citenamefont {Oettel}\ \emph {et~al.}(2012)\citenamefont {Oettel},
  \citenamefont {Dorosz}, \citenamefont {Berghoff}, \citenamefont {Nestler},\
  and\ \citenamefont {Schilling}}]{oettel_pre86_2012}%
  \BibitemOpen
  \bibfield  {author} {\bibinfo {author} {\bibfnamefont {M.}~\bibnamefont
  {Oettel}}, \bibinfo {author} {\bibfnamefont {S.}~\bibnamefont {Dorosz}},
  \bibinfo {author} {\bibfnamefont {M.}~\bibnamefont {Berghoff}}, \bibinfo
  {author} {\bibfnamefont {B.}~\bibnamefont {Nestler}}, \ and\ \bibinfo
  {author} {\bibfnamefont {T.}~\bibnamefont {Schilling}},\ }\href {\doibase
  10.1103/PhysRevE.86.021404} {\bibfield  {journal} {\bibinfo  {journal}
  {Phys.\ Rev.\ E}\ }\textbf {\bibinfo {volume} {86}},\ \bibinfo {pages}
  {021404} (\bibinfo {year} {2012})}\BibitemShut {NoStop}%
\bibitem [{\citenamefont {H\"artel}\ \emph {et~al.}(2012)\citenamefont
  {H\"artel}, \citenamefont {Oettel}, \citenamefont {Rozas}, \citenamefont
  {Egelhaaf}, \citenamefont {Horbach},\ and\ \citenamefont
  {L\"owen}}]{haertel_prl108_2012}%
  \BibitemOpen
  \bibfield  {author} {\bibinfo {author} {\bibfnamefont {A.}~\bibnamefont
  {H\"artel}}, \bibinfo {author} {\bibfnamefont {M.}~\bibnamefont {Oettel}},
  \bibinfo {author} {\bibfnamefont {R.~E.}\ \bibnamefont {Rozas}}, \bibinfo
  {author} {\bibfnamefont {S.~U.}\ \bibnamefont {Egelhaaf}}, \bibinfo {author}
  {\bibfnamefont {J.}~\bibnamefont {Horbach}}, \ and\ \bibinfo {author}
  {\bibfnamefont {H.}~\bibnamefont {L\"owen}},\ }\href {\doibase
  10.1103/PhysRevLett.108.226101} {\bibfield  {journal} {\bibinfo  {journal}
  {Phys.\ Rev.\ Lett.}\ }\textbf {\bibinfo {volume} {108}},\ \bibinfo {pages}
  {226101} (\bibinfo {year} {2012})}\BibitemShut {NoStop}%
\bibitem [{\citenamefont {H\"artel}\ \emph
  {et~al.}(2015{\natexlab{b}})\citenamefont {H\"artel}, \citenamefont {Kohl},\
  and\ \citenamefont {Schmiedeberg}}]{haertel_pre92_2015}%
  \BibitemOpen
  \bibfield  {author} {\bibinfo {author} {\bibfnamefont {A.}~\bibnamefont
  {H\"artel}}, \bibinfo {author} {\bibfnamefont {M.}~\bibnamefont {Kohl}}, \
  and\ \bibinfo {author} {\bibfnamefont {M.}~\bibnamefont {Schmiedeberg}},\
  }\href {\doibase 10.1103/PhysRevE.92.042310} {\bibfield  {journal} {\bibinfo
  {journal} {Phys. Rev. E}\ }\textbf {\bibinfo {volume} {92}},\ \bibinfo
  {pages} {042310} (\bibinfo {year} {2015}{\natexlab{b}})}\BibitemShut
  {NoStop}%
\bibitem [{\citenamefont {Alts}\ \emph {et~al.}(1987)\citenamefont {Alts},
  \citenamefont {Nielaba}, \citenamefont {D'Aguanno},\ and\ \citenamefont
  {Forstmann}}]{alts_cp111_1987}%
  \BibitemOpen
  \bibfield  {author} {\bibinfo {author} {\bibfnamefont {T.}~\bibnamefont
  {Alts}}, \bibinfo {author} {\bibfnamefont {P.}~\bibnamefont {Nielaba}},
  \bibinfo {author} {\bibfnamefont {B.}~\bibnamefont {D'Aguanno}}, \ and\
  \bibinfo {author} {\bibfnamefont {F.}~\bibnamefont {Forstmann}},\ }\href
  {\doibase 10.1016/0301-0104(87)80136-2} {\bibfield  {journal} {\bibinfo
  {journal} {Chem. Phys.}\ }\textbf {\bibinfo {volume} {111}},\ \bibinfo
  {pages} {223} (\bibinfo {year} {1987})}\BibitemShut {NoStop}%
\bibitem [{\citenamefont {Mier‐y‐Teran}\ \emph {et~al.}(1990)\citenamefont
  {Mier‐y‐Teran}, \citenamefont {Suh}, \citenamefont {White},\ and\
  \citenamefont {Davis}}]{mier-y-teran_jcp92_1990}%
  \BibitemOpen
  \bibfield  {author} {\bibinfo {author} {\bibfnamefont {L.}~\bibnamefont
  {Mier‐y‐Teran}}, \bibinfo {author} {\bibfnamefont {S.~H.}\ \bibnamefont
  {Suh}}, \bibinfo {author} {\bibfnamefont {H.~S.}\ \bibnamefont {White}}, \
  and\ \bibinfo {author} {\bibfnamefont {H.~T.}\ \bibnamefont {Davis}},\ }\href
  {\doibase http://dx.doi.org/10.1063/1.458542} {\bibfield  {journal} {\bibinfo
   {journal} {J. Chem. Phys.}\ }\textbf {\bibinfo {volume} {92}},\ \bibinfo
  {pages} {5087} (\bibinfo {year} {1990})}\BibitemShut {NoStop}%
\bibitem [{\citenamefont {Patra}\ and\ \citenamefont
  {Ghosh}(1993)}]{patra_pre47_1993}%
  \BibitemOpen
  \bibfield  {author} {\bibinfo {author} {\bibfnamefont {C.~N.}\ \bibnamefont
  {Patra}}\ and\ \bibinfo {author} {\bibfnamefont {S.~K.}\ \bibnamefont
  {Ghosh}},\ }\href {\doibase 10.1103/PhysRevE.47.4088} {\bibfield  {journal}
  {\bibinfo  {journal} {Phys. Rev. E}\ }\textbf {\bibinfo {volume} {47}},\
  \bibinfo {pages} {4088} (\bibinfo {year} {1993})}\BibitemShut {NoStop}%
\bibitem [{\citenamefont {Biben}\ \emph {et~al.}(1998)\citenamefont {Biben},
  \citenamefont {Hansen},\ and\ \citenamefont {Rosenfeld}}]{biben_pre57_1998}%
  \BibitemOpen
  \bibfield  {author} {\bibinfo {author} {\bibfnamefont {T.}~\bibnamefont
  {Biben}}, \bibinfo {author} {\bibfnamefont {J.~P.}\ \bibnamefont {Hansen}}, \
  and\ \bibinfo {author} {\bibfnamefont {Y.}~\bibnamefont {Rosenfeld}},\
  }\href@noop {} {\bibfield  {journal} {\bibinfo  {journal} {Phys. Rev. E}\
  }\textbf {\bibinfo {volume} {57}},\ \bibinfo {pages} {R3727} (\bibinfo {year}
  {1998})}\BibitemShut {NoStop}%
\bibitem [{\citenamefont {Gillespie}\ \emph {et~al.}(2003)\citenamefont
  {Gillespie}, \citenamefont {Nonner},\ and\ \citenamefont
  {Eisenberg}}]{gillespie_pre68_2003}%
  \BibitemOpen
  \bibfield  {author} {\bibinfo {author} {\bibfnamefont {D.}~\bibnamefont
  {Gillespie}}, \bibinfo {author} {\bibfnamefont {W.}~\bibnamefont {Nonner}}, \
  and\ \bibinfo {author} {\bibfnamefont {R.~S.}\ \bibnamefont {Eisenberg}},\
  }\href {\doibase 10.1103/PhysRevE.68.031503} {\bibfield  {journal} {\bibinfo
  {journal} {Phys. Rev. E}\ }\textbf {\bibinfo {volume} {68}},\ \bibinfo
  {pages} {031503} (\bibinfo {year} {2003})}\BibitemShut {NoStop}%
\bibitem [{\citenamefont {Forsman}\ \emph {et~al.}(2011)\citenamefont
  {Forsman}, \citenamefont {Woodward},\ and\ \citenamefont
  {Trulsson}}]{forsman_jpcb115_2011}%
  \BibitemOpen
  \bibfield  {author} {\bibinfo {author} {\bibfnamefont {J.}~\bibnamefont
  {Forsman}}, \bibinfo {author} {\bibfnamefont {C.~E.}\ \bibnamefont
  {Woodward}}, \ and\ \bibinfo {author} {\bibfnamefont {M.}~\bibnamefont
  {Trulsson}},\ }\href {\doibase 10.1021/jp111747w} {\bibfield  {journal}
  {\bibinfo  {journal} {J. Phys. Chem. B}\ }\textbf {\bibinfo {volume} {115}},\
  \bibinfo {pages} {4606} (\bibinfo {year} {2011})}\BibitemShut {NoStop}%
\bibitem [{\citenamefont {Wang}\ \emph {et~al.}(2011)\citenamefont {Wang},
  \citenamefont {Liu},\ and\ \citenamefont {Neretnieks}}]{wang_jpcm23_2011}%
  \BibitemOpen
  \bibfield  {author} {\bibinfo {author} {\bibfnamefont {Z.}~\bibnamefont
  {Wang}}, \bibinfo {author} {\bibfnamefont {L.}~\bibnamefont {Liu}}, \ and\
  \bibinfo {author} {\bibfnamefont {I.}~\bibnamefont {Neretnieks}},\ }\href
  {\doibase 10.1088/0953-8984/23/17/175002} {\bibfield  {journal} {\bibinfo
  {journal} {J. Phys.: Condens. Matter}\ }\textbf {\bibinfo {volume} {23}},\
  \bibinfo {pages} {175002} (\bibinfo {year} {2011})}\BibitemShut {NoStop}%
\bibitem [{\citenamefont {Waisman}\ and\ \citenamefont
  {Lebowitz}(1970)}]{waisman_jcp52_1970}%
  \BibitemOpen
  \bibfield  {author} {\bibinfo {author} {\bibfnamefont {E.}~\bibnamefont
  {Waisman}}\ and\ \bibinfo {author} {\bibfnamefont {J.~L.}\ \bibnamefont
  {Lebowitz}},\ }\href {\doibase 10.1063/1.1673642} {\bibfield  {journal}
  {\bibinfo  {journal} {J. Chem. Phys.}\ }\textbf {\bibinfo {volume} {52}},\
  \bibinfo {pages} {4307} (\bibinfo {year} {1970})}\BibitemShut {NoStop}%
\bibitem [{\citenamefont {H\o{}ye}\ \emph {et~al.}(1974)\citenamefont
  {H\o{}ye}, \citenamefont {Lebowitz},\ and\ \citenamefont
  {Stell}}]{hoye_jcp61_1974}%
  \BibitemOpen
  \bibfield  {author} {\bibinfo {author} {\bibfnamefont {J.~S.}\ \bibnamefont
  {H\o{}ye}}, \bibinfo {author} {\bibfnamefont {J.~L.}\ \bibnamefont
  {Lebowitz}}, \ and\ \bibinfo {author} {\bibfnamefont {G.}~\bibnamefont
  {Stell}},\ }\href {\doibase http://dx.doi.org/10.1063/1.1682485} {\bibfield
  {journal} {\bibinfo  {journal} {J. Chem. Phys.}\ }\textbf {\bibinfo {volume}
  {61}},\ \bibinfo {pages} {3253} (\bibinfo {year} {1974})}\BibitemShut
  {NoStop}%
\bibitem [{\citenamefont {Blum}(1975)}]{blum_mp30_1975}%
  \BibitemOpen
  \bibfield  {author} {\bibinfo {author} {\bibfnamefont {L.}~\bibnamefont
  {Blum}},\ }\href {\doibase 10.1080/00268977500103051} {\bibfield  {journal}
  {\bibinfo  {journal} {Mol. Phys.}\ }\textbf {\bibinfo {volume} {30}},\
  \bibinfo {pages} {1529} (\bibinfo {year} {1975})}\BibitemShut {NoStop}%
\bibitem [{\citenamefont {Stell}\ and\ \citenamefont
  {Sun}(1975)}]{stell_jcp63_1975}%
  \BibitemOpen
  \bibfield  {author} {\bibinfo {author} {\bibfnamefont {G.}~\bibnamefont
  {Stell}}\ and\ \bibinfo {author} {\bibfnamefont {S.~F.}\ \bibnamefont
  {Sun}},\ }\href {\doibase http://dx.doi.org/10.1063/1.431338} {\bibfield
  {journal} {\bibinfo  {journal} {J. Chem. Phys.}\ }\textbf {\bibinfo {volume}
  {63}},\ \bibinfo {pages} {5333} (\bibinfo {year} {1975})}\BibitemShut
  {NoStop}%
\bibitem [{\citenamefont {S\'anchez-D\'iaz}\ \emph {et~al.}(2010)\citenamefont
  {S\'anchez-D\'iaz}, \citenamefont {Vizcarra-Rend\'on},\ and\ \citenamefont
  {Medina-Noyola}}]{sanchez-diaz_jcp132_2010}%
  \BibitemOpen
  \bibfield  {author} {\bibinfo {author} {\bibfnamefont {L.~E.}\ \bibnamefont
  {S\'anchez-D\'iaz}}, \bibinfo {author} {\bibfnamefont {A.}~\bibnamefont
  {Vizcarra-Rend\'on}}, \ and\ \bibinfo {author} {\bibfnamefont
  {M.}~\bibnamefont {Medina-Noyola}},\ }\href {\doibase
  http://dx.doi.org/10.1063/1.3455336} {\bibfield  {journal} {\bibinfo
  {journal} {J. Chem. Phys.}\ }\textbf {\bibinfo {volume} {132}},\ \bibinfo
  {pages} {234506} (\bibinfo {year} {2010})}\BibitemShut {NoStop}%
\bibitem [{\citenamefont {Henderson}\ \emph {et~al.}(1979)\citenamefont
  {Henderson}, \citenamefont {Blum},\ and\ \citenamefont
  {Lebowitz}}]{henderson_jecie102_1979}%
  \BibitemOpen
  \bibfield  {author} {\bibinfo {author} {\bibfnamefont {D.}~\bibnamefont
  {Henderson}}, \bibinfo {author} {\bibfnamefont {L.}~\bibnamefont {Blum}}, \
  and\ \bibinfo {author} {\bibfnamefont {J.~L.}\ \bibnamefont {Lebowitz}},\
  }\href {\doibase http://dx.doi.org/10.1016/S0022-0728(79)80459-3} {\bibfield
  {journal} {\bibinfo  {journal} {J. Electroanal. Chem.}\ }\textbf {\bibinfo
  {volume} {102}},\ \bibinfo {pages} {315} (\bibinfo {year}
  {1979})}\BibitemShut {NoStop}%
\bibitem [{\citenamefont {Gillespie}(2014)}]{gillespie_pre90_2014}%
  \BibitemOpen
  \bibfield  {author} {\bibinfo {author} {\bibfnamefont {D.}~\bibnamefont
  {Gillespie}},\ }\href {\doibase 10.1103/PhysRevE.90.052134} {\bibfield
  {journal} {\bibinfo  {journal} {Phys. Rev. E}\ }\textbf {\bibinfo {volume}
  {90}},\ \bibinfo {pages} {052134} (\bibinfo {year} {2014})}\BibitemShut
  {NoStop}%
\bibitem [{\citenamefont {Ulander}\ and\ \citenamefont
  {Kjellander}(2001)}]{ulander_jcp114_2001}%
  \BibitemOpen
  \bibfield  {author} {\bibinfo {author} {\bibfnamefont {J.}~\bibnamefont
  {Ulander}}\ and\ \bibinfo {author} {\bibfnamefont {R.}~\bibnamefont
  {Kjellander}},\ }\href {\doibase 10.1063/1.1350449} {\bibfield  {journal}
  {\bibinfo  {journal} {J. Chem. Phys.}\ }\textbf {\bibinfo {volume} {114}},\
  \bibinfo {pages} {4893} (\bibinfo {year} {2001})}\BibitemShut {NoStop}%
\bibitem [{\citenamefont {Evans}\ and\ \citenamefont
  {Henderson}(2009)}]{evans_jpcm21_2009}%
  \BibitemOpen
  \bibfield  {author} {\bibinfo {author} {\bibfnamefont {R.}~\bibnamefont
  {Evans}}\ and\ \bibinfo {author} {\bibfnamefont {J.~R.}\ \bibnamefont
  {Henderson}},\ }\href {\doibase 10.1088/0953-8984/21/47/474220} {\bibfield
  {journal} {\bibinfo  {journal} {J. Phys.: Condens. Matter}\ }\textbf
  {\bibinfo {volume} {21}},\ \bibinfo {pages} {474220} (\bibinfo {year}
  {2009})}\BibitemShut {NoStop}%
\bibitem [{\citenamefont {Hansen-Goos}\ and\ \citenamefont
  {Roth}(2006)}]{hansen-goos_jpcm18_2006}%
  \BibitemOpen
  \bibfield  {author} {\bibinfo {author} {\bibfnamefont {H.}~\bibnamefont
  {Hansen-Goos}}\ and\ \bibinfo {author} {\bibfnamefont {R.}~\bibnamefont
  {Roth}},\ }\href {\doibase 10.1088/0953-8984/18/37/002} {\bibfield  {journal}
  {\bibinfo  {journal} {J.\ Phys.: Condens.\ Matter}\ }\textbf {\bibinfo
  {volume} {18}},\ \bibinfo {pages} {8413} (\bibinfo {year}
  {2006})}\BibitemShut {NoStop}%
\bibitem [{\citenamefont {H{\"a}rtel}\ \emph {et~al.}(2015)\citenamefont
  {H{\"a}rtel}, \citenamefont {Janssen}, \citenamefont {Samin},\ and\
  \citenamefont {van Roij}}]{haertel_jpcm27_2015}%
  \BibitemOpen
  \bibfield  {author} {\bibinfo {author} {\bibfnamefont {A.}~\bibnamefont
  {H{\"a}rtel}}, \bibinfo {author} {\bibfnamefont {M.}~\bibnamefont {Janssen}},
  \bibinfo {author} {\bibfnamefont {S.}~\bibnamefont {Samin}}, \ and\ \bibinfo
  {author} {\bibfnamefont {R.}~\bibnamefont {van Roij}},\ }\href {\doibase
  doi:10.1088/0953-8984/27/19/194129} {\bibfield  {journal} {\bibinfo
  {journal} {J. Phys.: Condens. Matter}\ }\textbf {\bibinfo {volume} {27}},\
  \bibinfo {pages} {194129} (\bibinfo {year} {2015})}\BibitemShut {NoStop}%
\bibitem [{\citenamefont {Tarazona}\ \emph {et~al.}(1985)\citenamefont
  {Tarazona}, \citenamefont {Evans},\ and\ \citenamefont
  {Marconi}}]{tarazona_mph54_1985}%
  \BibitemOpen
  \bibfield  {author} {\bibinfo {author} {\bibfnamefont {P.}~\bibnamefont
  {Tarazona}}, \bibinfo {author} {\bibfnamefont {R.}~\bibnamefont {Evans}}, \
  and\ \bibinfo {author} {\bibfnamefont {U.~M.~B.}\ \bibnamefont {Marconi}},\
  }\href {\doibase 10.1080/00268978500101051} {\bibfield  {journal} {\bibinfo
  {journal} {Mol.\ Phys.}\ }\textbf {\bibinfo {volume} {54}},\ \bibinfo {pages}
  {1357} (\bibinfo {year} {1985})}\BibitemShut {NoStop}%
\bibitem [{\citenamefont {Tarazona}(2000)}]{tarazona_prl84_2000}%
  \BibitemOpen
  \bibfield  {author} {\bibinfo {author} {\bibfnamefont {P.}~\bibnamefont
  {Tarazona}},\ }\href {\doibase 10.1103/PhysRevLett.84.694} {\bibfield
  {journal} {\bibinfo  {journal} {Phys.\ Rev.\ Lett.}\ }\textbf {\bibinfo
  {volume} {84}},\ \bibinfo {pages} {694} (\bibinfo {year} {2000})}\BibitemShut
  {NoStop}%
\bibitem [{\citenamefont {Oettel}\ \emph {et~al.}(2010)\citenamefont {Oettel},
  \citenamefont {G\"orig}, \citenamefont {H\"artel}, \citenamefont {L\"owen},
  \citenamefont {Radu},\ and\ \citenamefont {Schilling}}]{oettel_pre82_2010}%
  \BibitemOpen
  \bibfield  {author} {\bibinfo {author} {\bibfnamefont {M.}~\bibnamefont
  {Oettel}}, \bibinfo {author} {\bibfnamefont {S.}~\bibnamefont {G\"orig}},
  \bibinfo {author} {\bibfnamefont {A.}~\bibnamefont {H\"artel}}, \bibinfo
  {author} {\bibfnamefont {H.}~\bibnamefont {L\"owen}}, \bibinfo {author}
  {\bibfnamefont {M.}~\bibnamefont {Radu}}, \ and\ \bibinfo {author}
  {\bibfnamefont {T.}~\bibnamefont {Schilling}},\ }\href {\doibase
  10.1103/PhysRevE.82.051404} {\bibfield  {journal} {\bibinfo  {journal}
  {Phys.\ Rev.\ E}\ }\textbf {\bibinfo {volume} {82}},\ \bibinfo {pages}
  {051404} (\bibinfo {year} {2010})}\BibitemShut {NoStop}%
\bibitem [{\citenamefont {Merlet}\ \emph {et~al.}(2011)\citenamefont {Merlet},
  \citenamefont {Salanne}, \citenamefont {Rotenberg},\ and\ \citenamefont
  {Madden}}]{merlet_jpcc115_2011}%
  \BibitemOpen
  \bibfield  {author} {\bibinfo {author} {\bibfnamefont {C.}~\bibnamefont
  {Merlet}}, \bibinfo {author} {\bibfnamefont {M.}~\bibnamefont {Salanne}},
  \bibinfo {author} {\bibfnamefont {B.}~\bibnamefont {Rotenberg}}, \ and\
  \bibinfo {author} {\bibfnamefont {P.~A.}\ \bibnamefont {Madden}},\ }\href
  {\doibase 10.1021/jp205461g} {\bibfield  {journal} {\bibinfo  {journal} {J.
  Phys. Chem. C}\ }\textbf {\bibinfo {volume} {115}},\ \bibinfo {pages} {16613}
  (\bibinfo {year} {2011})}\BibitemShut {NoStop}%
\bibitem [{\citenamefont {Vega}\ \emph {et~al.}(2003)\citenamefont {Vega},
  \citenamefont {Abascal},\ and\ \citenamefont {McBride}}]{vega_jcp119_2003}%
  \BibitemOpen
  \bibfield  {author} {\bibinfo {author} {\bibfnamefont {C.}~\bibnamefont
  {Vega}}, \bibinfo {author} {\bibfnamefont {J.~L.~F.}\ \bibnamefont
  {Abascal}}, \ and\ \bibinfo {author} {\bibfnamefont {C.}~\bibnamefont
  {McBride}},\ }\href {\doibase 10.1063/1.1576374} {\bibfield  {journal}
  {\bibinfo  {journal} {J. Chem. Phys.}\ }\textbf {\bibinfo {volume} {119}},\
  \bibinfo {pages} {964} (\bibinfo {year} {2003})}\BibitemShut {NoStop}%
\bibitem [{\citenamefont {Hynninen}\ \emph {et~al.}(2006)\citenamefont
  {Hynninen}, \citenamefont {Leunissen}, \citenamefont {van Blaaderen},\ and\
  \citenamefont {Dijkstra}}]{hynninen_prl96_2006}%
  \BibitemOpen
  \bibfield  {author} {\bibinfo {author} {\bibfnamefont {A.-P.}\ \bibnamefont
  {Hynninen}}, \bibinfo {author} {\bibfnamefont {M.~E.}\ \bibnamefont
  {Leunissen}}, \bibinfo {author} {\bibfnamefont {A.}~\bibnamefont {van
  Blaaderen}}, \ and\ \bibinfo {author} {\bibfnamefont {M.}~\bibnamefont
  {Dijkstra}},\ }\href {\doibase 10.1103/PhysRevLett.96.018303} {\bibfield
  {journal} {\bibinfo  {journal} {Phys. Rev. Lett.}\ }\textbf {\bibinfo
  {volume} {96}},\ \bibinfo {pages} {018303} (\bibinfo {year}
  {2006})}\BibitemShut {NoStop}%
\bibitem [{\citenamefont {Chaumont}\ \emph {et~al.}(2005)\citenamefont
  {Chaumont}, \citenamefont {Schurhammer}, ,\ and\ \citenamefont
  {Wipff}}]{chaumont_jpcb109_2005}%
  \BibitemOpen
  \bibfield  {author} {\bibinfo {author} {\bibfnamefont {A.}~\bibnamefont
  {Chaumont}}, \bibinfo {author} {\bibfnamefont {R.}~\bibnamefont
  {Schurhammer}}, , \ and\ \bibinfo {author} {\bibfnamefont {G.}~\bibnamefont
  {Wipff}},\ }\href {\doibase 10.1021/jp052854h} {\bibfield  {journal}
  {\bibinfo  {journal} {J. Phys. Chem. B}\ }\textbf {\bibinfo {volume} {109}},\
  \bibinfo {pages} {18964} (\bibinfo {year} {2005})}\BibitemShut {NoStop}%
\bibitem [{\citenamefont {Limbach}\ \emph {et~al.}(2006)\citenamefont
  {Limbach}, \citenamefont {Arnold}, \citenamefont {Mann},\ and\ \citenamefont
  {Holm}}]{espresso}%
  \BibitemOpen
  \bibfield  {author} {\bibinfo {author} {\bibfnamefont {H.-J.}\ \bibnamefont
  {Limbach}}, \bibinfo {author} {\bibfnamefont {A.}~\bibnamefont {Arnold}},
  \bibinfo {author} {\bibfnamefont {B.~A.}\ \bibnamefont {Mann}}, \ and\
  \bibinfo {author} {\bibfnamefont {C.}~\bibnamefont {Holm}},\ }\href {\doibase
  10.1016/j.cpc.2005.10.005} {\bibfield  {journal} {\bibinfo  {journal}
  {Comput. Phys. Commun.}\ }\textbf {\bibinfo {volume} {174}},\ \bibinfo
  {pages} {704} (\bibinfo {year} {2006})}\BibitemShut {NoStop}%
\bibitem [{\citenamefont {Mezger}\ \emph {et~al.}(2015)\citenamefont {Mezger},
  \citenamefont {Roth}, \citenamefont {Schröder}, \citenamefont {Reichert},
  \citenamefont {Pontoni},\ and\ \citenamefont
  {Reichert}}]{mezger_jcp142_2015}%
  \BibitemOpen
  \bibfield  {author} {\bibinfo {author} {\bibfnamefont {M.}~\bibnamefont
  {Mezger}}, \bibinfo {author} {\bibfnamefont {R.}~\bibnamefont {Roth}},
  \bibinfo {author} {\bibfnamefont {H.}~\bibnamefont {Schröder}}, \bibinfo
  {author} {\bibfnamefont {P.}~\bibnamefont {Reichert}}, \bibinfo {author}
  {\bibfnamefont {D.}~\bibnamefont {Pontoni}}, \ and\ \bibinfo {author}
  {\bibfnamefont {H.}~\bibnamefont {Reichert}},\ }\href {\doibase
  http://dx.doi.org/10.1063/1.4918742} {\bibfield  {journal} {\bibinfo
  {journal} {J. Chem. Phys.}\ }\textbf {\bibinfo {volume} {142}},\ \bibinfo
  {pages} {164707} (\bibinfo {year} {2015})}\BibitemShut {NoStop}%
\bibitem [{\citenamefont {Rica}\ \emph {et~al.}(2012)\citenamefont {Rica},
  \citenamefont {Ziano}, \citenamefont {Salerno}, \citenamefont {Mantegazza},\
  and\ \citenamefont {Brogioli}}]{rica_prl109_2012}%
  \BibitemOpen
  \bibfield  {author} {\bibinfo {author} {\bibfnamefont {R.~A.}\ \bibnamefont
  {Rica}}, \bibinfo {author} {\bibfnamefont {R.}~\bibnamefont {Ziano}},
  \bibinfo {author} {\bibfnamefont {D.}~\bibnamefont {Salerno}}, \bibinfo
  {author} {\bibfnamefont {F.}~\bibnamefont {Mantegazza}}, \ and\ \bibinfo
  {author} {\bibfnamefont {D.}~\bibnamefont {Brogioli}},\ }\href {\doibase
  10.1103/PhysRevLett.109.156103} {\bibfield  {journal} {\bibinfo  {journal}
  {Phys. Rev. Lett.}\ }\textbf {\bibinfo {volume} {109}},\ \bibinfo {pages}
  {156103} (\bibinfo {year} {2012})}\BibitemShut {NoStop}%
\end{thebibliography}
\end{document}